\documentclass[manuscript]{aastex}
\usepackage{epsfig}
\usepackage{color}

\slugcomment{Accepted for publication in the {\it Astronomical Journal}}

\shorttitle{Chandra Observation of NGC 7793}
\shortauthors{Pannuti et al.}

\begin{document}


\title{A {\it Chandra} Observation of the Nearby Sculptor Group Sd Galaxy NGC 7793}


\author{Thomas G. Pannuti\altaffilmark{1}, Eric M. Schlegel\altaffilmark{2}, 
Miroslav D. Filipovi\'{c}\altaffilmark{3}, Jeffrey L. Payne\altaffilmark{3}, Robert Petre\altaffilmark{4},
Ilana M. Harrus\altaffilmark{5},  Wayne D. Staggs\altaffilmark{1} and Christina K. 
Lacey\altaffilmark{6}}


\altaffiltext{1}{Space Science Center, Department of Earth and Space Sciences, Morehead 
State University, 235 Martindale Drive, Morehead, KY 40351; t.pannuti@moreheadstate.edu,
wdstag01@moreheadstate.edu}
\altaffiltext{2}{Department of Physics and Astronomy, 
University of Texas-San Antonio, 6900 North Loop 1604 West, San Antonio, TX,
78249-0663; eric.schlegel@utsa.edu}
\altaffiltext{3}{School of Computing and Mathematics, 
University of Western Sydney, Locked Bag 1797, Penrith South, DC,
New South Wales, 1797, Australia; m.filipovic@uws.edu.au, astronomer@me.com}
\altaffiltext{4}{NASA/GSFC, X-ray Astrophysics Laboratory, Code 662, Greenbelt, MD, 
20771; robert.petre-1@nasa.gov}
\altaffiltext{5}{NASA/HQ, Astrophysics Division, MS 3W39, Washington, DC 20546-0001;
Ilana.M.Harrus@nasa.gov}
\altaffiltext{6}{Department of Physics and Astronomy, Hofstra University, 102 Berliner
Hall, Hempstead, NY 11549; Christina.Lacey@hofstra.edu}


\begin{abstract}

  We conducted a {\it Chandra} ACIS observation of the nearby Sculptor
  Group Sd galaxy NGC 7793 as part of a multiwavelength study of
  supernova remnants (SNRs) in nearby galaxies.  At the assumed
  distance to NGC 7793 of 3.91 Mpc, the limiting unabsorbed luminosity
  of the detected discrete X-ray sources is $L$$_X$ (0.2 -- 10.0 keV)
  $\approx$ 3$\times$10$^{36}$ ergs s$^{-1}$.  A total of 22 discrete
  sources were detected at the $\sim$3$\sigma$ level or greater
  including one ultra-luminous X-ray source (ULX).  Based on
  multiwavelength comparisons, we identify X-ray sources coincident with 
  one SNR, the candidate microquasar N7793-S26, one HII region and 
  two foreground Galactic stars.  We also find that the X-ray counterpart
  to the candidate radio SNR R3 is time-variable in its X-ray emission: we
  therefore rule out the possibility that this source is a single SNR. A
  marked asymmetry is seen in the distribution of the discrete sources
  with the majority lying in the eastern half of this galaxy. All of
  the sources were analyzed using quantiles to estimate spectral
  properties and spectra of the four brightest sources (including the
  ULX) were extracted and analyzed.  We searched for time-variability
  in the X-ray emission of the detected discrete sources using our
  measured fluxes along with fluxes measured from prior {\it Einstein}
  and {\it ROSAT} observations. From this study, three discrete X-ray
  sources are
  established to be significantly variable.  A spectral analysis of
  the galaxy's diffuse emission is characterized by a temperature of
  ${\it kT}$ = 0.19-0.25 keV.  The luminosity function of the
  discrete sources shows a slope with an absolute value of $\Gamma$ = 
  $-$0.65$\pm$0.11 if we exclude the ULX. If the ULX is included, the
  luminosity function has a long tail to high L$_X$ with a
  poor-fitting slope of $\Gamma$ = $-$0.62$\pm$0.2.  The ULX-less slope is
  comparable to the slopes measured for the distributions of NGC 6946
  and NGC 2403 but much shallower than the slopes measured for the
  distributions of IC 5332 and M83. Lastly, we comment on the multi-wavelength 
  properties of the SNR population of NGC 7793.

\end{abstract}






\section{Introduction}

Based on its superior angular resolution capabilities -- namely, an on-axis point 
spread function (PSF) with a half-power diameter of $\sim$1$\arcsec$ -- the 
{\it Chandra} X-ray Observatory \citep{Weisskopf02} is an ideal instrument
for surveying populations of discrete X-ray sources in nearby spiral
galaxies. To date, numerous nearby spiral galaxies have been
the subjects of deep {\it Chandra} observations which have sampled
their resident X-ray source populations in unprecedented detail. Prominent
examples of such galaxies which have been the subject of such studies 
include M33 \citep{Plucinsky08,Long10,Tuellmann11}, M51 \citep{Terashima01, 
Terashima04}, M81 
\citep{Swartz03}, M83 \citep{Soria03}, M101 \citep{Pence01, Mukai03}, NGC 1637
\citep{Immler03}, NGC 2403 \citep{Schlegel03A}, NGC 3184 \citep{Kilgard02}, 
NGC 6946 \citep{Holt03} and IC 5332 \citep{Kilgard02}. In each case, the 
{\it Chandra} observations have dramatically increased the numbers of known 
discrete X-ray sources in each galaxy. The applications of these observations 
include sampling a robust number of discrete X-ray sources for
statistically-significant analyses, spatially resolving the discrete sources
from a component of diffuse X-ray emission detected from some galaxies, 
spectral analyses of the brightest discrete sources, time variability analyses 
(often incorporating observations made with previous X-ray observatories) and 
measurement with high
accuracy of the positions of X-ray sources for the purposes of identifying
counterparts at other wavelengths.

\par

Typically, the classes of X-ray objects detected by these surveys
include X-ray sources associated with active galactic nuclei
(AGN), X-ray binaries (XRBs) and supernova remnants
(SNRs). Observations of XRBs and SNRs are essential in developing a
thorough understanding of stellar evolution.  Unfortunately, studies
of XRBs and SNRs in our own Galaxy are hampered by observational
difficulties,  including significant absorption along Galactic lines of
sight as well as considerable uncertainties in distances to these sources. In
addition, the Galactic population of XRBs and SNRs represent a galaxy of
a single mass, metallicity, star formation history and morphological type.
Observing XRBs 
and SNRs located in nearby galaxies minimizes these issues.
In prior works \citep{Pannuti00, Lacey01, Pannuti02, Pannuti07, Filipovic08}, 
we analyzed high angular resolution observations at multiple wavelengths of 
several nearby galaxies to both identify SNRs and statistically assess their
properties.  In the present paper we continue this work by analyzing a
{\it Chandra} observation of the nearby Sd galaxy NGC 7793. This
observation was conducted primarily to study X-ray emission from the
SNR population in NGC 7793; here we consider the properties of the
discrete X-ray sources detected in this galaxy as well as the
accompanying diffuse X-ray emission.

\par

NGC 7793, a member of the nearby Sculptor Group \citep{Puche88}, lies
at a distance of 3.91 Mpc \citep{Karachentsev03} and at an 
inclination angle of $i$$\sim$50$^{\circ}$
\citep{Tully88}.  General properties of both NGC 7793 and the pointed
{\it Chandra} observation of this galaxy are listed in Table
\ref{Table1}.  NGC 7793 has been the subject of prior X-ray
observations made with the {\it Einstein} Imaging Proportional Counter
(IPC) \citep{Fabbiano92} and the {\it R\"{o}ntgensatellit}  ({\it ROSAT}) Position 
Sensitive Proportional Counter (PSPC) (\citealt{Read99}, hereafter RP99).  These
observations detected seven X-ray sources -- including an
ultra-luminous X-ray source (ULX) located along the southern edge of
the galaxy -- within the optical extent of NGC 7793.  In addition,
prominent diffuse X-ray emission which permeates much of the disk of
the galaxy has been detected (RP99). The SNR population of NGC 7793
has been well-studied by both optical and radio surveys
\citep{Blair97,Pannuti02}; based on these searches, a total of 32
resident SNRs have been identified in this galaxy. In the present
paper, we will concentrate on searching for X-ray counterparts to these 32
sources. N7793-S26 -- an additional 
source that was initially classified as a SNR by \citet{Blair97} and
detected in the radio by \citet{Read99} and \citet{Pannuti02} -- has recently
been classified as a microquasar candidate by \citet{Pakull10} and \citet{Soria10}. We exclude
this source from our investigation presented here of the properties of SNRs in 
NGC 7793 and we will discuss this source (particularly its X-ray, optical and
radio properties when compared to extragalactic superbubbles) in more detail 
in a future paper.

\par

The organization of this paper is as follows: the observations and
data reduction are described in Section \ref{ObsSection}. Properties
of the discrete X-ray sources detected by this observation --
including spectral analysis of four of the most luminous sources as well
as searches for multi-wavelength counterparts and time-variable
emission from all sources -- are discussed in Section
\ref{DiscreteSection}. Next we discuss the diffuse X-ray emission from
NGC 7793 as sampled by this observation (Section
\ref{DiffuseSection}), the luminosity function of the discrete sources
(Section \ref{LFSection}) and the properties of the SNR population of
this galaxy (Section \ref{SNRSection}). Finally, we summarize our
results in Section \ref{ConclusionsSection}.


\section{Observations and Data Reduction\label{ObsSection}}


We used the Advanced Charge-Coupled Device (CCD) Imaging Spectrometer
(ACIS) \citep{Gar2003} onboard {\it Chandra} to observe NGC 7793. The
observation was obtained in Very Faint mode on 2003 September 6-7
using an aim point approximately two arcminutes west of the nucleus,
ensuring the maximum coverage of NGC 7793 across the
(back-illuminated) ACIS-S3 chip.  The exposure lasted approximately
49724 seconds and, after correcting for the deadtime, the effective
exposure time was 49094 seconds.  \par We accumulated source-free
background areas offset from the galaxy (namely from the
back-illuminated ACIS-S1 chip) and extracted a light curve using
50-second bins to test for the presence of soft background flares.  No
flaring behavior of any kind was detected.  We re-filtered the Level 1
data, correcting for the induced charge-transfer inefficiency
following the prescription of \cite{TB2000}.  This approach permits
using a single event redistribution matrix in the spectral fitting,
altering the response matrix for the off-axis-dependent effective
area. The data was reduced with standard tools in the software application package  
``{\it Chandra} Interactive Analysis of Observations" ({\it CIAO}\footnote{See
http://cxc.harvard.edu/ciao/ and \citet{Fruscione06}.}) Version 3.4 (CALDB version 3.3.0).
\par 
Point sources were identified using {\tt wavdetect} at
1$''$, 2$''$, and 4$''$ scales \citep{Freeman02}.  The detected
sources were merged into a final source list after eliminating
duplicate detections.  Source counts were extracted using apertures
that increased with off-axis angle to ensure the inclusion of an
approximately constant fraction of the PSF.
The minimum aperture was 2$''$ in diameter and enclosed $>$95\% of the
PSF.  These same apertures were used to extract
spectra (see Section \ref{LuminousXraySourcesSection}) or counts for
quantiles (see Section \ref{QuantileSection}), depending upon the count
rate.  A background spectrum was obtained from a region on the ACIS-S3
chip but outside of the galaxy and southeast of the nucleus.

\section{The Discrete X-ray Source Population\label{DiscreteSection}}

\begin{figure}
\includegraphics[angle=0]{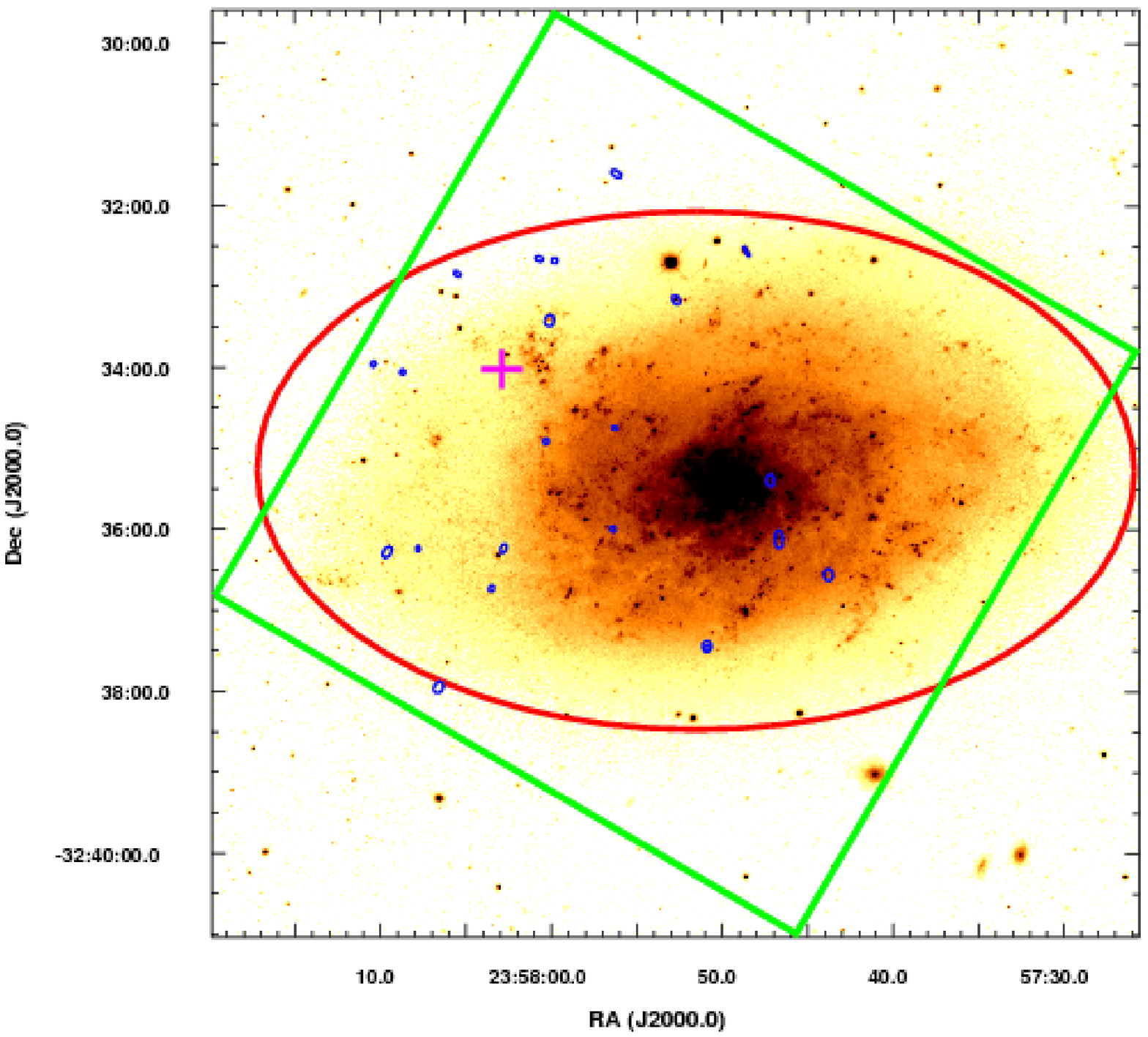}
\caption{R-band image of NGC 7793 with the positions of
the detected sources (corresponding to 90\% confidence levels) indicated
with blue ellipses. The green box indicates the field of view of the ACIS-S3 chip 
during the observation, the red ellipse indicates the optical extent of NGC 7793 and
the magenta cross indicates the aimpoint of the observation.
The sizes of the source location ellipses are related to the errors in the 
determination of the source positions. Notice the asymmetric distribution in the positions
of the sources. See Section \ref{DiscreteSection} for a discussion about
the discrete sources and Section \ref{AsymmetrySubSection} for a discussion
about the asymmetric distribution.\label{fig1}}
\end{figure}

We detected 22 discrete X-ray sources within the optical extent of NGC 7793 
at the $\sim$3$\sigma$ level or greater  to a limiting unabsorbed luminosity of 
approximately $L$$_X$ $\approx$ 3 $\times$ 10$^{36}$ ergs s$^{-1}$ over the 
energy range of 0.2 
through 10.0 keV, assuming a foreground column density of $N$$_H$ =
1.15$\times$10$^{20}$ cm$^{-2}$ and a power law model with a photon
index $\Gamma$=1.5.  Table \ref{Table2} lists the properties of these
sources, including position (in J2000.0 coordinates), absorbed and
unabsorbed fluxes, unabsorbed luminosities and the
significance (in $\sigma$) of the detection of the source.  In Figure
\ref{fig1}, we present an R-band image\footnote{The R-band image has
been kindly provided to us by Annette M. N. Ferguson: for details on
the observations of this galaxy that produced this image, the reader
is referred to \citet{Ferguson96}.} of NGC 7793 with the positions of
the detected X-ray sources indicated with ellipses representing the
90\% confidence contours of their measured positions.  The sizes of these
ellipses are related to the errors in the determination of the source positions. 
In comparison,
the {\it ROSAT} PSPC revealed seven discrete sources (RP99).  Similar
to the previous X-ray observations (\citealt{Fabbiano92}, RP99), we
have not detected a central X-ray source associated with the nucleus
of this galaxy to the stated limiting unabsorbed luminosity.  We expect that the
majority of the detected sources are resident XRBs; other
possibilities include background AGNs, foreground stars, and X-ray
luminous SNRs. Table \ref{Table3} lists identified counterparts to the X-ray sources as 
detected at multiple wavelengths: we discuss each of these associations
in detail in Sections \ref{LuminousXraySourcesSection} and 
\ref{OtherXraySourcesSection}. Lastly, Table \ref{Table4} contains the 
spectral properties of the
entire population of discrete X-ray sources based on quantiles; these
properties will be discussed in more detail in Section \ref{QuantileSection}.

\par

Inspection of Figure 1 reveals a remarkable asymmetry in the
distribution of the discrete X-ray sources with the large majority of
the discrete sources located in the eastern half of NGC 7793. Only
four of the 22 discrete X-ray sources are found in the western half
and there is a stark absence of discrete sources in the northwestern
quadrant.  Previous {\it Einstein} and {\it ROSAT} observations
(\citealt{Fabbiano92}, RP99) also did not detect any discrete sources
in this quadrant.  We discuss this asymmetry more thoroughly in Section
\ref{AsymmetrySubSection}.

\subsection{The Most Luminous Discrete Sources\label{LuminousXraySourcesSection}}

For four of the 22 discrete X-ray sources, there were sufficient
counts above our arbitrary limit of 200 counts (corresponding to a count rate of
$\sim$4$\times$10$^{-3}$ counts per second) to extract spectra and generate
spectral fits that yielded parameters which were measured to $<$30\%.  We
now discuss general properties of each of these four sources: properties of the 
other discrete sources are presented in Section \ref{OtherXraySourcesSection}.
In each case, the spectra were fit using the software package {\it
XSPEC} Version 11.3.1 \citep{Arnaud96}\footnote{Also see
http://heasarc.gsfc.nasa.gov/docs/xanadu/xspec/.}: the individual
spectra were grouped to a minimum of 25 counts per bin and each
spectrum was fit over the energy range where there were a sufficient
number of counts. We have not performed any fits to the X-ray spectrum of
the source CXOU J235752.7$-$323309 even though the number of counts
detected from this source exceed the threshold stated above because this source
is physically associated with a foreground star (see Section \ref{OtherXraySourcesSection}).
\par
We used four basic models to fit each extracted spectrum: a simple
power law model, a bremsstrahlung model \citep{Karzas61, Kellogg75},
the APEC model, an optically thin thermal plasma model known as
the APEC model \citep{Smith01}, and
finally the DiskBB model \citep{Mitsuda84, Makishima86}. This last
model describes the spectrum from an accretion disk consisting of
multiple blackbody components and is characterized by the temperature
$T$$_{in}$ at the inner disk radius. To account for photoelectric
absorption along the line of sight we used the Wisconsin cross-section
models \citep{Morrison83}.  There are two choices possible when
fitting the column density: fixing it at the value of the known
Galactic column density in the pointing direction (that is, 
$N$$_{\rm  H}$ = 1.15$\times$10$^{20}$ cm$^{-2}$) or treating it as a free
parameter.  With this in mind, we performed fits to the extracted
spectra using the four models first with the column density frozen to
the known Galactic column and then with the column density left as a
free parameter.  Lastly, a background spectrum was extracted using an
aperture 0.7 arcmin in diameter and positioned off the optical extent
of the galaxy.  No point sources were included in this aperture.  The
resulting spectrum extracted from this aperture was accurately fit
using a combination of a power law component and Gaussians. In Table \ref{Table4},
we present a representative summary of our derived best fits and not the results
of every fit that we attempted obtained for each model. The
background spectrum itself and best-fit model were then included in
the fits to each point source without adjusting the background model
fit components.
\par
In general, the basic models return
statistically acceptable fits for each source. In some cases, a
particular spectrum warranted a more complex model to obtain a
satisfactory fit. Also, in some cases significant statistical
differences in the fits were seen when the column density was frozen
or thawed: below we discuss specific results for fitting the extracted
spectra.  The spectra of two of these sources were also analyzed by
RP99: we also compare our derived fits with the fits obtained by those
authors.

\par

{\it CXOU J235746.7$-$323607}: This source corresponds to the X-ray
source P10 identified by RP99 and the candidate radio SNR R3
identified by \citet{Pannuti02}: the offset between CXOU
J235746.7$-$323607 and P10 is $\sim$6$\arcsec$ while the offset
between CXOU J235746.7$-$323607 and R3 is only $\sim$1.5$\arcsec$.
\citet{Pannuti02} proposed that this X-ray source and the candidate
radio SNR are physically associated based on their positional
proximity.  RP99 commented on the soft nature of the {\it ROSAT}
PSPC spectrum of this source.  Based on this soft spectrum, a lack of apparent
variability in the X-ray emission (between two epochs of {\it ROSAT}
observations) and its positional association with
a portion of NGC 7793 that features numerous SNRs, RP99
speculated that this X-ray source may be a superbubble or the
collective X-ray emission from multiple SNRs.
\par
Our fits indicate a moderately-hard spectrum for this source.  The
bremsstrahlung temperature is higher than that of RP99 (${\it kT}$
$\sim$ 1.6 keV compared to ${\it kT}$ $\sim$ 0.8 keV), although the
error bars overlap the fitted values.  The fitted temperature obtained from
the DiskBB model is softer (${\it kT}$ $\sim$ 0.7 keV): the use of the DiskBB 
model in this case -- while physically less appropriate for a source identified 
as an SNR, is justified below.  Within the errors, the fitted column
density is consistent with the known column toward NGC 7793.  If we
fix the column density to the known column, the impact is largely
confined to the power law model: the power law index ${\Gamma}$ falls
from 2.9 to 1.7 with a slight overlap in the error bars; for the other
models, the fitted parameters differ by $<$10\%.  All models (with the exception 
of the APEC model) require the addition of two zero-width Gaussians to model line-like
features.  The Gaussians have line centers at 0.94 and 1.46 keV and
the equivalent widths of these lines are largest in the power law
model, with values of 111 and 102 eV, respectively. 
\par
\begin{figure}
\includegraphics[angle=-90,scale=0.75]{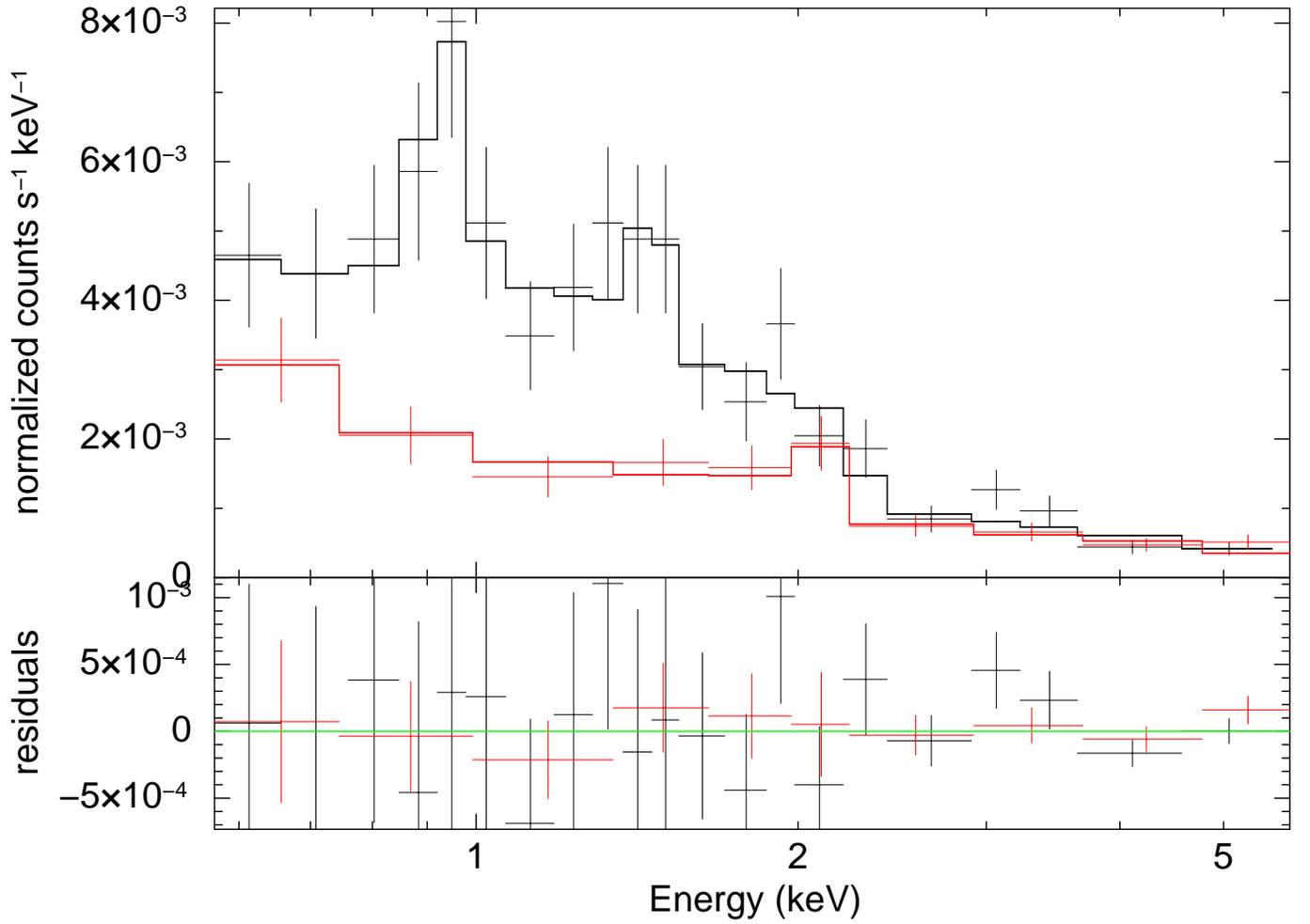}
\caption{Extracted spectrum of CXOU J235746.7$-$323607
as fit with a power law model (column density thawed). The source
spectrum is shown in black while the background spectrum is shown
in red. \label{fig2}}
\end{figure}
\begin{figure}
\includegraphics[angle=-90,scale=0.75]{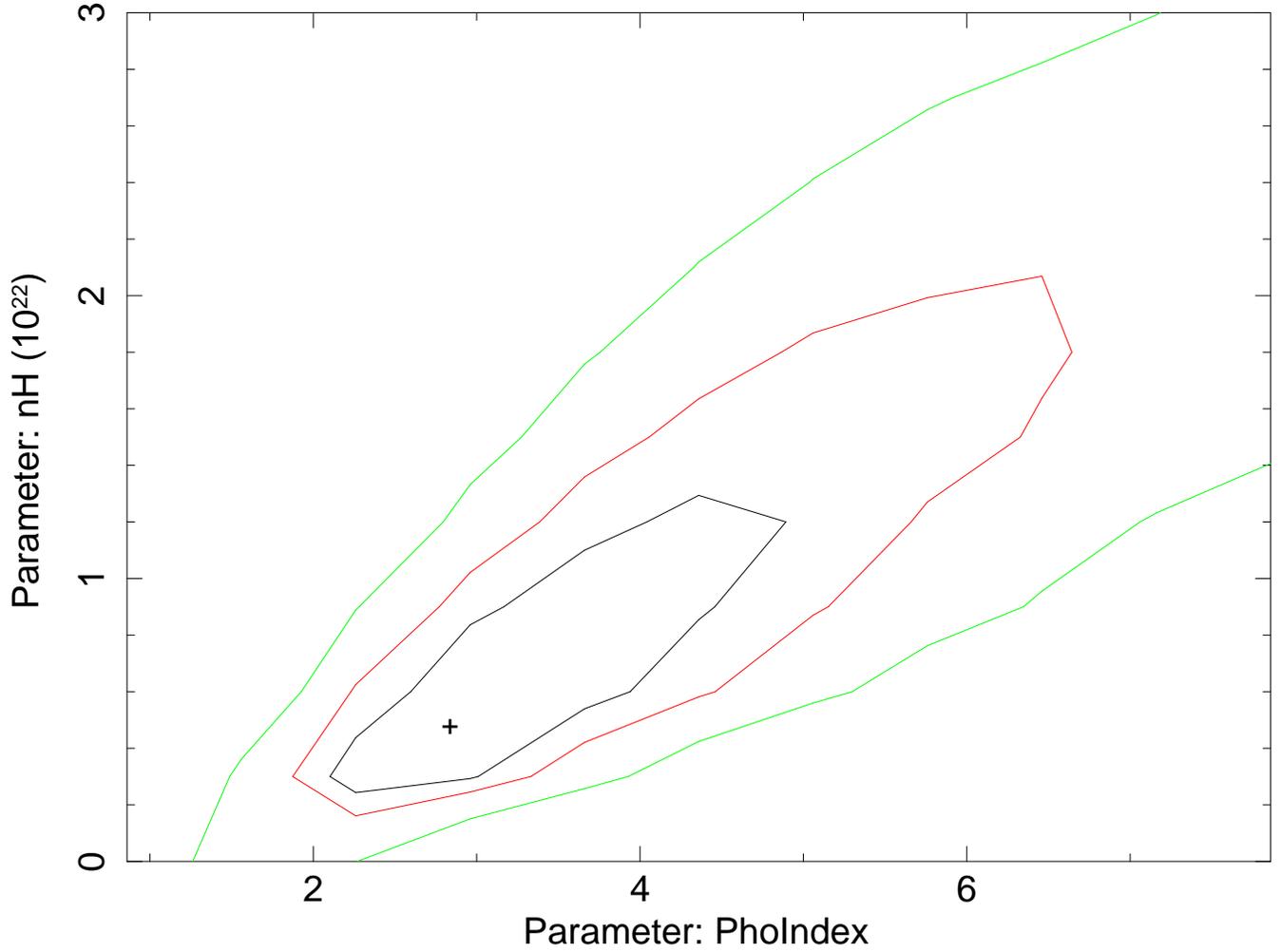}
\caption{Confidence contours for fit (as shown in
Figure \ref{fig2}) to extracted spectrum of CXOU 
J235746.7$-$323607. The contours are at the 1$\sigma$, 2$\sigma$ and
3$\sigma$ levels.\label{fig3}}
\end{figure}

The normalization of the DiskBB model is defined as 
[($R$$_{\rm in}$/km)/($D$/10 kpc)]$^2$ where $R$$_{\rm in}$ is the inner 
disk radius in km and $D$ is the distance to the source in units of 10 kpc.  The fitted
normalization then corresponds to 26$^{+73}_{-17}$ km, a value larger
than is generally deemed typical of neutron stars but falling within
inferred radii of other low-mass X-ray binaries (e.g.,
\citet{Church01}). In Figures \ref{fig2} and \ref{fig3} we present the extracted
spectrum of CXOU J235746.7$-$323607 as fit with the power law model
(with a variable column density) and a confidence contour plot for this
fit, respectively. 
\par
We note that we have detected a clear variability in the
X-ray emission from this source. The prior {\it ROSAT} PSPC
observations of this galaxy reported by RP99 caught this source in a
brighter state than our {\it Chandra} observation by a factor of
approximately three. This time-variability -- coupled with the high X-ray
luminosity and the moderately hard spectrum observed for this source -- 
cast doubt on the classification of this source as a single SNR. Alternatively, 
CXOU J235746.7$-$323607 may be an SNR/XRB system analogous to the 
Galactic source W50/SS 433 \citep{SafiHarb99}, though the observed X-ray
luminosity of the former source is several orders of magnitudes greater than
the latter source. It is possible that the observed X-ray emission stems from 
a complex of sources which remain unresolved even with the high angular
resolution capabilities of {\it Chandra}. The classification of CXOU J235746.7$-$323607
is therefore currently uncertain and we will discuss this source again in Section
\ref{VariabilitySection} when we discuss a search for time-variability in the X-ray
emission from the detected discrete X-ray sources and in Section
\ref{SNRSection} when we describe properties of the SNR population of
this galaxy.

\par

{\it CXOU J235750.9$-$323726}: This source, suspected to be a ULX
associated with NGC 7793, was identified as P13 by RP99. Those authors
presented a detailed history and an analysis of its spectral properties. To
summarize, this source was first detected by {\it Einstein}
\citep{Fabbiano92} and subsequently \citet{Margon85} included it in an
atlas of X-ray selected quasi-stellar objects, arguing that the
source was associated with a background quasar seen just below the
southern edge of NGC 7793. This quasar has been cataloged as 2355$-$329 
and features a redshift of 0.071. It has also been cataloged as 2355$-$3254 
in more recent observations presented by \citet{Bowen94}. The {\it Einstein}
observation localized the position to within an arcminute; with the
better angular resolution of the {\it ROSAT} PSPC, RP99 ruled out an
association between P13 and the background quasar, arguing instead
that the X-ray source is native to NGC 7793. RP99 described time
variability in the source's emission by comparing observations made
six months apart and speculated that the source may be either a
background galaxy or a black hole X-ray binary with an estimated mass
of $\sim$ 10 $M$$_{\odot}$. The estimated X-ray luminosity of this
source ($\sim$ 10$^{39}$ ergs s$^{-1}$, assuming that it is in fact
associated with NGC 7793) is comparable to ULXs seen in other
galaxies.

\par 

Our {\it Chandra} observations verify that this source is indeed
located within the optical extent of NGC 7793.  We used the improved
positional accuracy to search for a counterpart using optical
(H$\alpha$ and R-band) images and our radio maps of NGC 7793
\citep{Pannuti02} but we do not find a clear optical or radio
counterpart. Recently, \citet{Motch11} identified an optical
counterpart (a $V$ $\sim$ 20.5 magnitude star) and suggest this star
(a late B-type supergiant with a mass between 10 and 20 $M$$_{\odot}$)
to be the companion star to the observed X-ray source.

\par

\begin{figure}
\epsscale{0.90}
\includegraphics[angle=-90,scale=0.75]{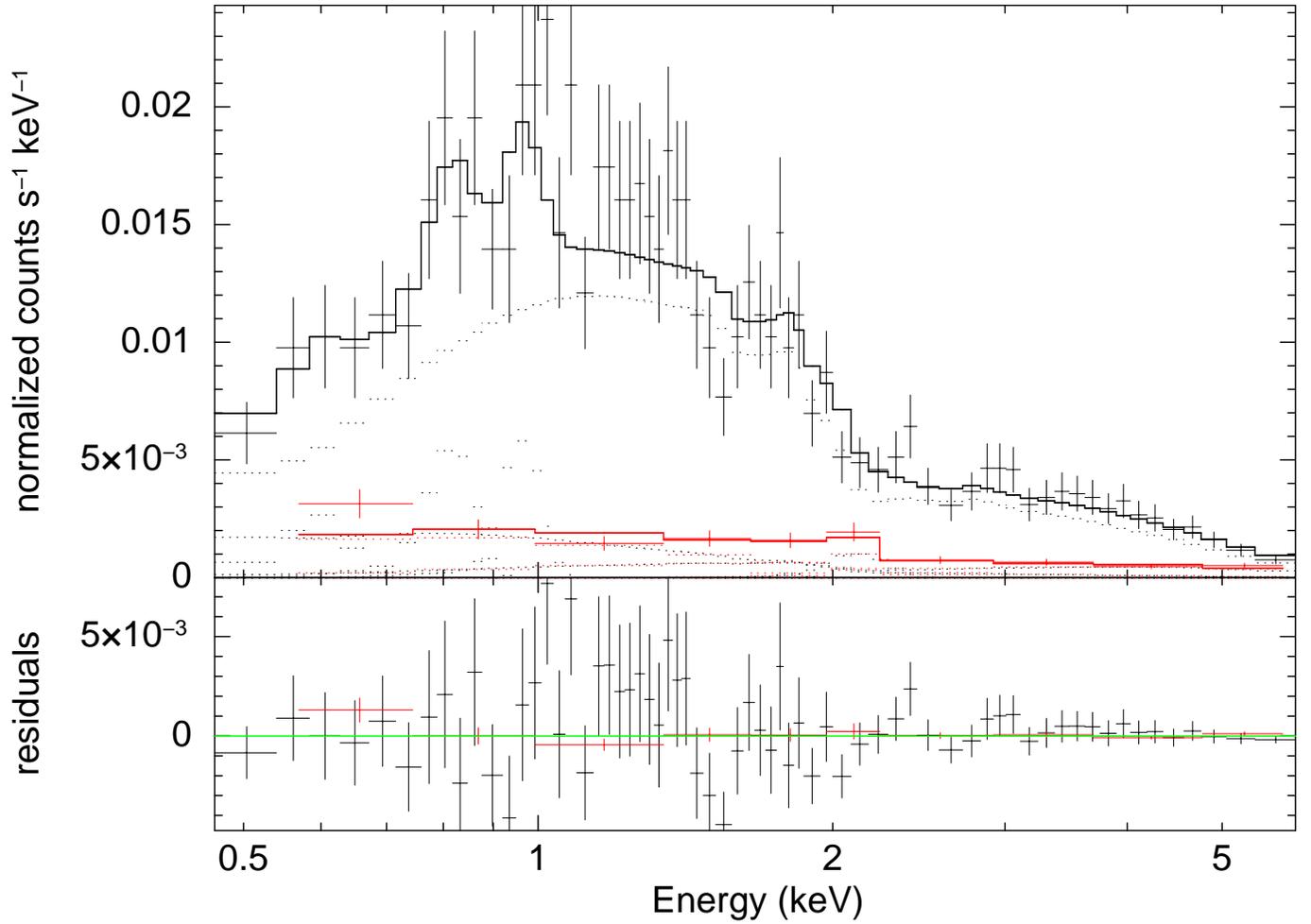}
\caption{Extracted spectrum of CXOU J235750.9$-$323726
as fit with the DiskBB model (column density thawed). The source
spectrum is shown in black while the background spectrum is shown
in red. Dashed lines show contributions of the different model components.\label{fig4}}
\end{figure}

\begin{figure}
\includegraphics[angle=-90,scale=0.75]{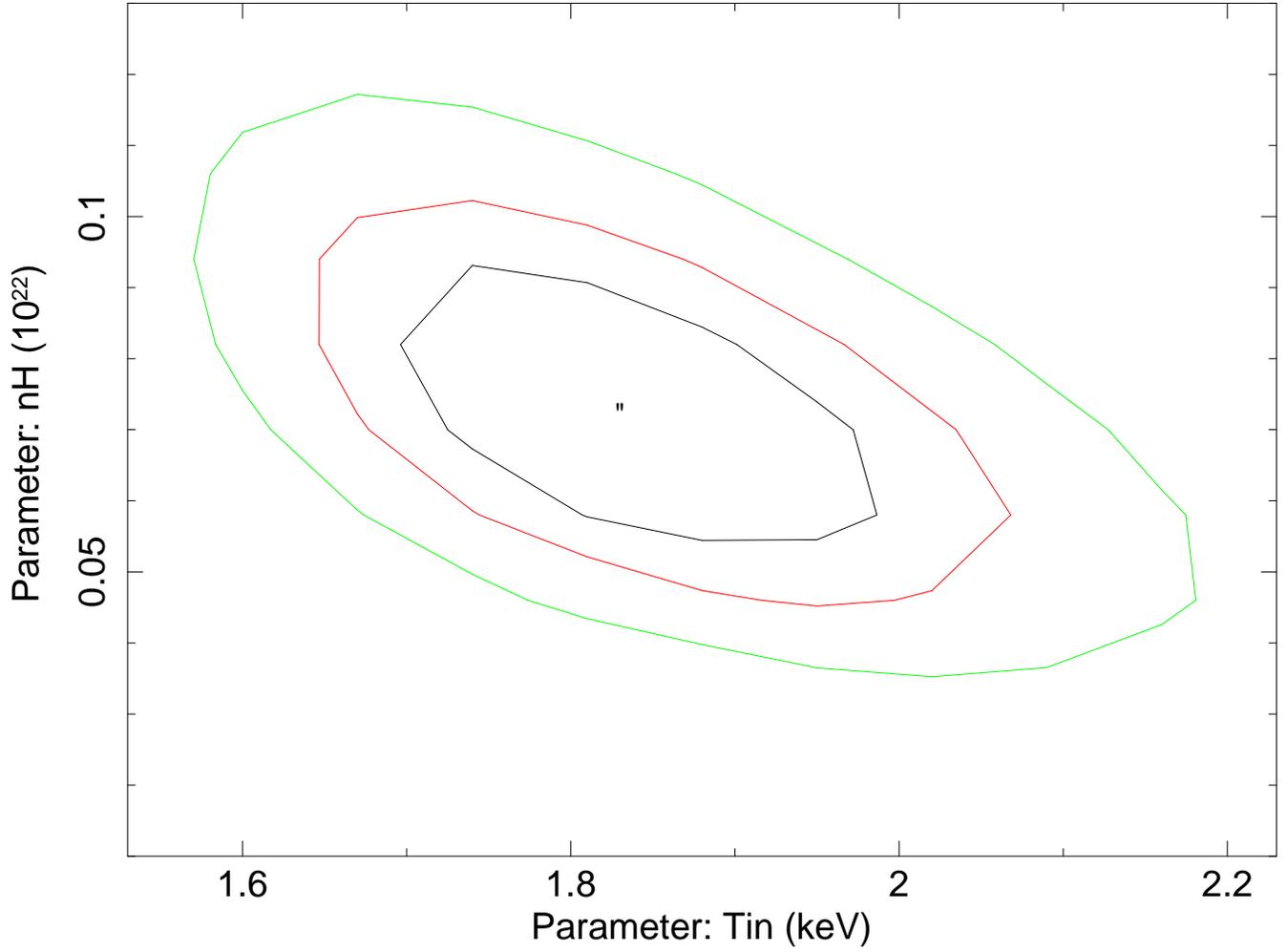}
\caption{Confidence contours for fit (as shown in Figure \ref{fig4}) 
to extracted spectrum of CXOU J235750.9$-$323726. The contours are
plotted at the same levels as in Figure \ref{fig3}.\label{fig5}}
\end{figure}

RP99 derived fits to their extracted {\it ROSAT} PSPC spectrum of this
X-ray source using either a thermal bremsstrahlung model with a
characteristic temperature ${\it kT}$ = 3.49$^{+4.26}_{-3.49}$ keV or
a power law model with a photon index $\Gamma$ $\sim$ 1.8$\pm$0.5.
We do not derive a statistically-acceptable fit for any models if the
column density is fixed to the Galactic value. The derived column 
densities from our fits were $N$$_H$ $\sim$ 10$^{21}$ cm$^{-2}$, nearly
a full order of magnitude greater than the nominal column density
toward NGC 7793 itself. 
Compared with the fit presented by RP99, our fit with a bremsstrahlung model 
returns a significantly higher
effective temperature ($kT$ $>$ 14 keV). A portion of the discrepancy
may be explained by the broader energy range sampled by the {\it
Chandra} spectrum; alternatively, a spectral state change is also
possible. The photon index derived by our power law fit
($\Gamma$=1.4$^{+0.20}_{-0.18}$) is consistent with the value derived
by RP99.  From the DiskBB model, our derived inner disk radius
temperature of $kT$$_{in}$ = 1.83$^{+0.24}_{-0.18}$ keV is consistent
with a stellar-mass black hole XRB.  The DiskBB model
yields a low value for the column density ($N$$_{\rm H}$ =
0.09$^{+0.02}_{-0.05}$ $\times$10$^{22}$ cm$^{-2}$): this value is
consistent with the column values of RP99. Figures \ref{fig4} and \ref{fig5}
show the extracted spectrum of CXOU J235750.9$-$323726 as fit with the
DiskBB model and a confidence contour plot for this fit, respectively.

\par

We also note that {\it Chandra} has revealed for the first time a
second source located $\sim$ 2$''$ east of the CXOU
J235750.9$-$323726. This second source is denoted as CXOU
J235750.9$-$323728 and it is an order of magnitude less luminous than
the ULX. CXOU J235750.9$-$323726 and CXOU J235750.9$-$323728 are
approximately 5$\arcsec$ and 8$\arcsec$ respectively from the position
given by RP99 for P13.  These sources would certainly be blended by
the broader PSF of the {\it ROSAT} PSPC
($\sim$ 27$\arcsec$).



\par

\begin{figure}
\epsscale{0.90}
\includegraphics[angle=-90,scale=0.75]{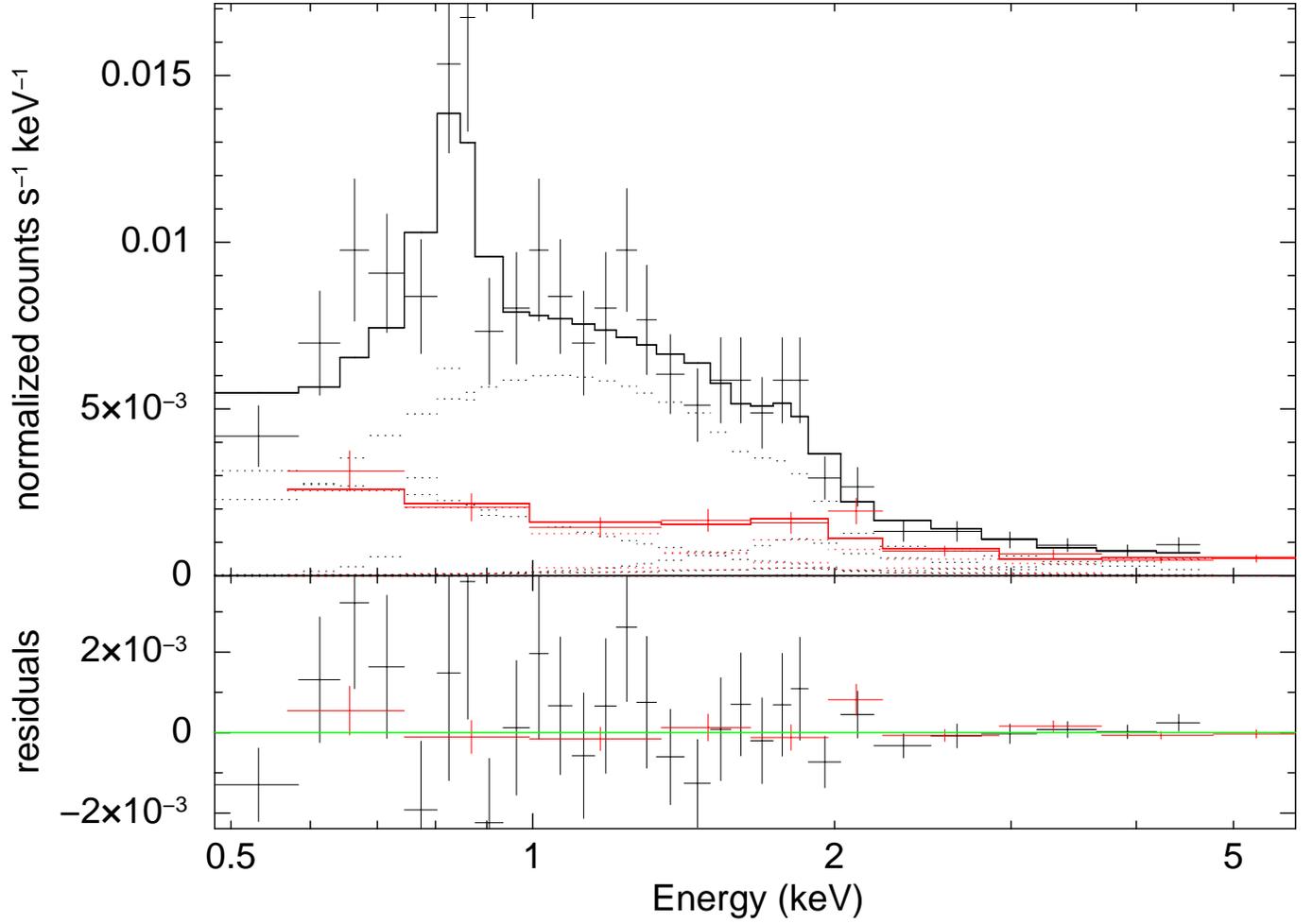}
\caption{Extracted spectrum of CXOU J235806.6$-$323757
as fit with a power law model. The source spectrum is shown in black while 
the background spectrum is shown in red. Dashed lines show contributions of 
the different model components.\label{fig6}}
\end{figure}

\begin{figure}
\includegraphics[angle=-90,scale=0.75]{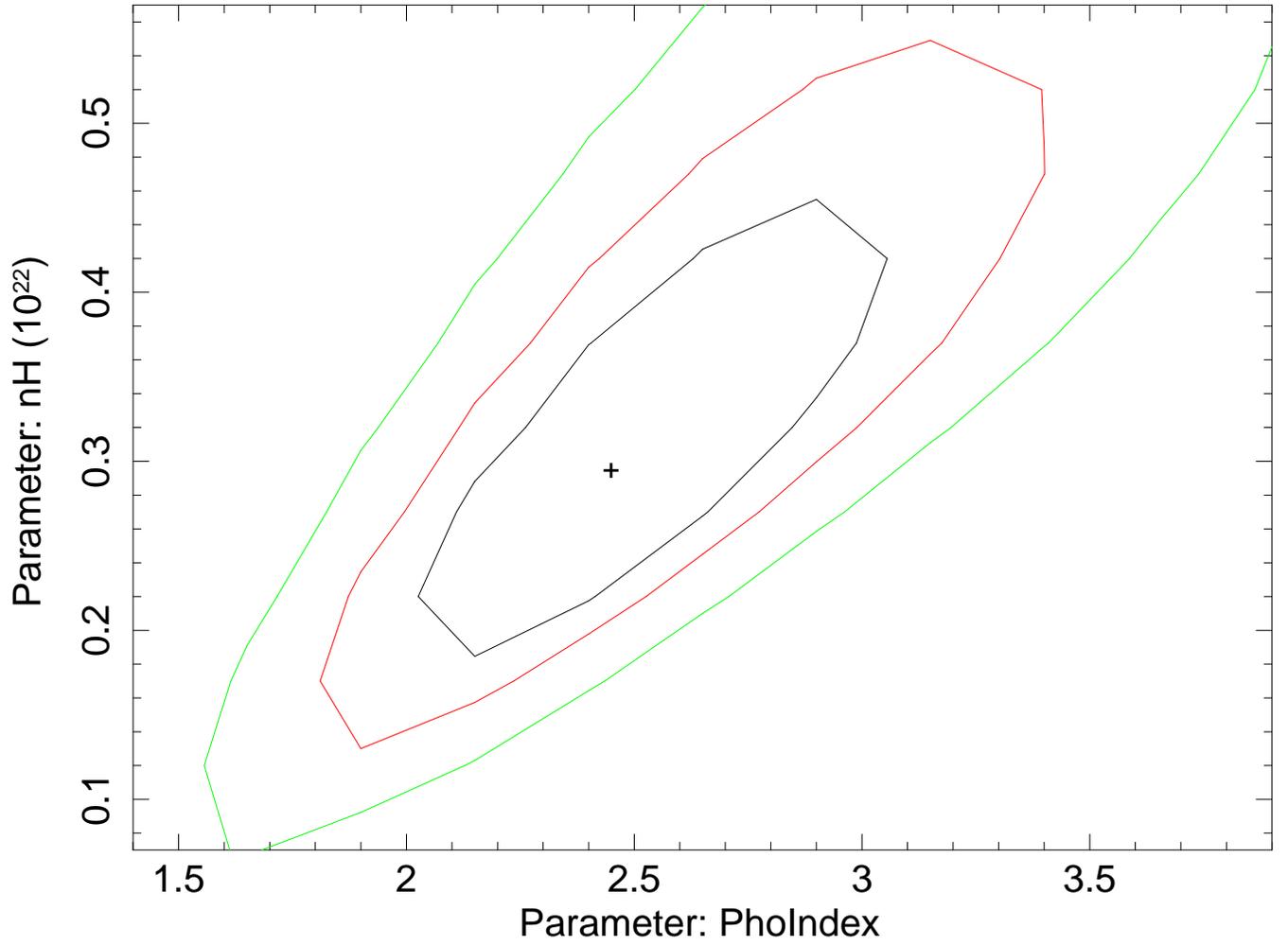}
\caption{Confidence contours for fit (as shown in Figure \ref{fig6}) 
to extracted spectrum of CXOU J235806.6$-$323757. The contours are
plotted at the same levels as in Figure \ref{fig3}.\label{fig7}}
\end{figure}

{\it CXOU J235806.6$-$323757}. The {\it Chandra} detection of this
source immediately establishes it as a variable -- given its measured
luminosity, it should have been detected during the prior {\it ROSAT}
observations.  The spectral models all return equally acceptable
fits: an order-of-magnitude higher column density than the known Galactic
value is required for fits with the bremsstrahlung model and the power law model
while the DiskBB model only requires a column of $N$$_H$$<$8$\times$10$^{20}$ 
cm$^{-2}$.
The bremsstrahlung temperature is $\it{kT}$ $\sim$2 keV but rises to $\it{kT}$ $\sim$ 
6.4 keV if the column density is fixed at the known value; otherwise, the
parameters of the fixed column models change insignificantly.  All
models require an unresolved line at 0.83 keV with equivalent widths that
range from 45 to 122 eV.  The bremsstrahlung and power law models require a
second unresolved line at 0.67 keV with an equivalent width of
$\sim$120 eV.  The combination of spectral fit values and variability
suggests an XRB classification is the most likely for this source.  Figures
\ref{fig6} and \ref{fig7} present the extracted spectrum of this
source as fit with the power law model (with a thawed column density)
and a confidence contour plot for this fit, respectively.

\par
\begin{figure}
\includegraphics[angle=-90,scale=0.70]{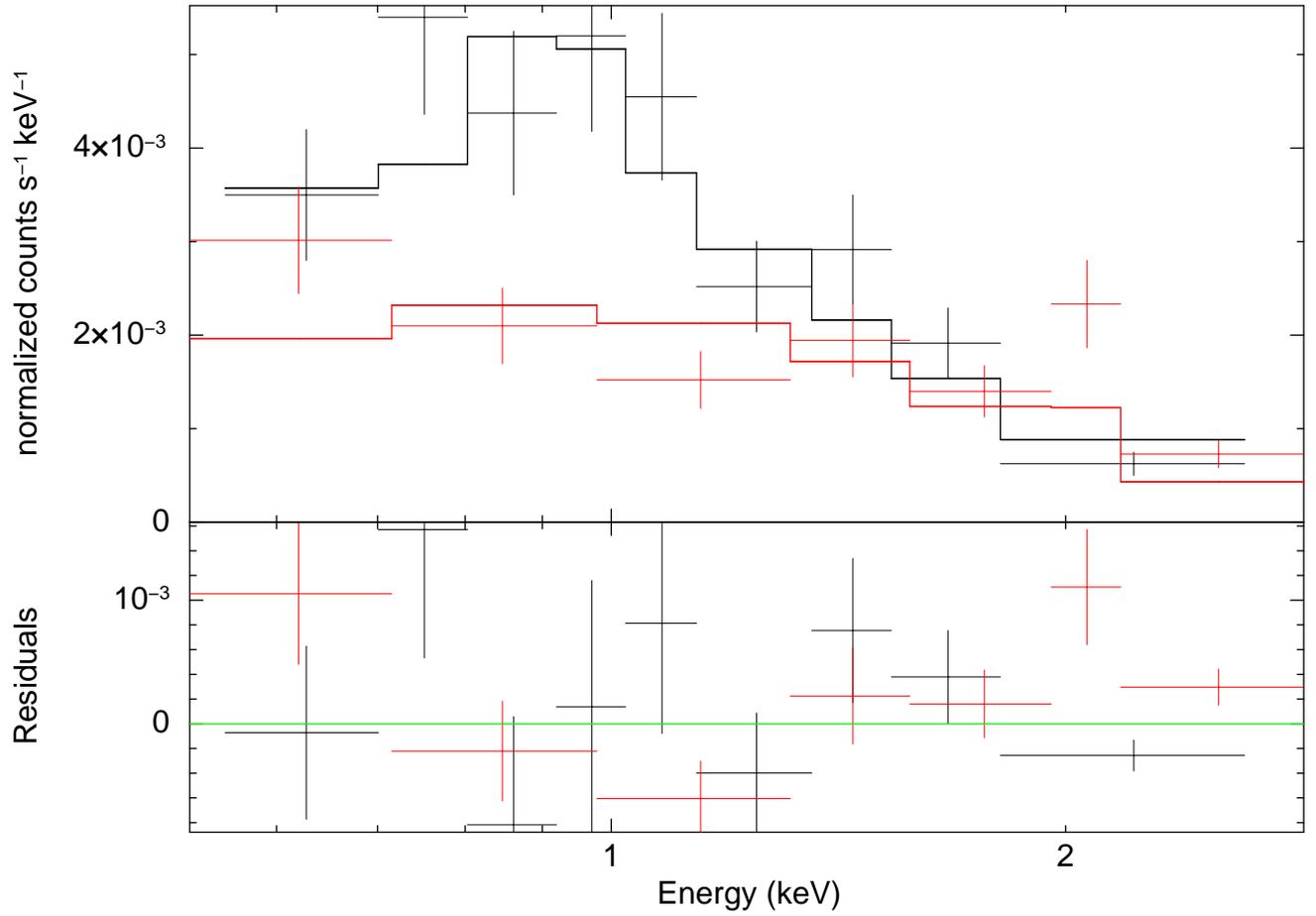}
\caption{Extracted spectrum of CXOU J235808.7$-$323403 as fit 
with the APEC model. The source spectrum is shown in black while 
the background spectrum is shown in red.\label{fig8}}
\end{figure}

{\it CXOU J235808.7$-$323403}. The offset between this source and RP99
source P9 is approximately $\sim$2$\arcsec$, so we claim that these
sources are in fact the same. This source is the weakest source of the four
considered in this Section for which a basic spectral analysis is 
reasonable. The three best-fit models are listed in Table~\ref{Table4}; none of them is
particularly good as single high (or low) bins contribute relatively large values to 
$\chi$$^2$.  We do not include other models as the fits are significantly 
poorer.  The spectrum is clearly soft for all three fits, which overlap within the errors.
Given the relatively poor fit, the flux is likely an overestimate and the errors on the 
normalization appropriately reflect the uncertainty. RP99 note that this source,
which corresponds to P9 in their paper, is variable and highly absorbed. If we 
adopt the best-fit APEC model, we confirm the high absorption as two of the fits
yield column densities with values of $N$$_H$ $\sim$ 0.1--0.2 $\times$10$^{22}$
cm$^{-2}$. As with RP99, we do not detect a counterpart at optical or radio
wavelengths. In Figure \ref{fig8} we present the extracted spectrum for this
source as fit with the APEC model. 



\subsection{Identifications of Other Discrete X-ray Sources\label{OtherXraySourcesSection}}

We now briefly comment on the nature of some of the other {\it Chandra}-detected
X-ray sources.  We searched at multiple wavelengths for counterparts
to these sources: we also identified sources which
may have been confused in prior X-ray observations due to poorer
angular resolution.  We recount the results of these searches here and present
a summary of these counterparts in Table \ref{Table3}.

\par

{\it Foreground Stars}: The optical counterparts of CXOU
J235748.6$-$323234 and CXOU J235752.7$-$323309 are foreground stars.
The first source corresponds to the star USNO 0574$-$1250312 \citep{Monet03}: it 
was previously detected with the {\it ROSAT} PSPC by RP99 and was labeled
by those authors as P6.  The offset between P6 and our {\it Chandra}
position is 4.2$\arcsec$; {\it Chandra}'s improved position yields an
offset from USNO 0574$-$1250312 of only 1.3$\arcsec$.
\citet{Davoust80} noted that the star USNO 0574$-$1250312 was
misclassified as an HII region (\#18) by \citet{Hodge69}.

The second source matches the source cataloged as P7 by RP99 with a
{\it Chandra}-{\it ROSAT} position offset of $\sim$2$\arcsec$.  We
therefore claim that the two sources are in fact the same. RP99
speculated that this source may be a background object; we instead
claim that this source is physically associated with the star USNO
0574$-$1250339 \citep{Monet03}, located only 0.8$\arcsec$ from the
{\it Chandra} position.

\par

{\it HII Regions}: Catalogs of HII regions in NGC 7793 have been
presented by \citet{Hodge69} and \citet{Davoust80}.  We used a search
radius of 3$\arcsec$ to identify associations between our {\it
Chandra} sources and the cataloged HII regions.  We found one such
association: the X-ray source CXOU J235743.8$-$323633 is offset from
the HII region D22 \citep{Davoust80} by 2.5$\arcsec$; we suggest that
these two sources are physically associated. This X-ray source may be
an SNR or an XRB associated with the HII region as we do not generally
expect H~II regions to be X-ray luminous.

\par

{\it SNRs}: As described previously, a total of 32
optically-identified SNRs and candidate radio SNRs have been
identified in NGC 7793 based on the optical (\citealt{Blair97}) and
radio searches (\citealt{Pannuti02}).  We have already discussed the
association between the X-ray source CXOU J235746.7$-$323607 and the
candidate radio SNR R3 (\S\ref{LuminousXraySourcesSection}). If we adopt
a search radius of 1.5$\arcsec$ (which corresponds to a linear
distance of approximately 30 pc at the assumed distance to NGC 7793), we 
find one other association\footnote{We note here that the published positions of
several candidate radio SNRs detected in NGC 7793 by \citet{Pannuti02}
are in error and give the correct positions here (RA units = hours,
minutes and seconds; Dec units are degrees, arcminutes and
arcseconds): NGC 7793-R1 -- RA (J2000.0): 23 57 40.2, Dec (J2000.0):
$-$32 36 38; NGC 7793-R4 -- RA (J2000.0): 23 57 48.2, Dec (J2000.0):
$-$32 36 15; NGC 7793-R5 -- RA (J2000.0): 23 58 00.6, Dec (J2000.0):
$-$32 35 06.}. This association (with an offset of 1.1$\arcsec$) occurs between
CXOU J235747.2$-$323523 and the optically-identified SNR S11
\citep{Blair97}. This source had been originally classified as an HII
region (\#40) by \citet{Davoust80}. \citet{Pannuti02} found a
non-thermal radio counterpart, which helped to solidify its
classification as an SNR.  The {\it Chandra} observation clearly
reveals an X-ray counterpart for the first time, making it one of a
small number of known extragalactic SNRs which have been detected at
X-ray, optical and radio wavelengths. We will discuss the
multi-wavelength properties of the SNR population of NGC 7793 in more
detail in Section \ref{SNRSection}. For completeness, we also note
that within this adopted search radius, 
our cataloged source CXOU J235800.1$-$323325 is coincident with the southern 
component of the candidate
microquasar N7793-S26.  As mentioned earlier, this source was previously classified
as an SNR \citep{Blair97,Pannuti02} but has recently been re-classified as a microquasar
\citep{Pakull10,Soria10}: CXOU J235800.1$-$323325 features some 
spatial extent in the X-ray that mimics the observed extended emission seen at
optical and radio wavelengths.

\par   

{\it Young Massive Star Clusters}: We searched for positional
coincidences between our sample of X-ray sources and the 20 young
massive star clusters identified in this galaxy by \citet{Larsen99A}
and \citet{Larsen99B}. The effective radii of these clusters were
given by \citet{Larsen99B} and range from $\sim$4-60 pc at our assumed
distance to NGC 7793 (corresponding to a projected angular scale of
0.3--3$\arcsec$).  Using these radii, our search found no positional matches.

\par

{\it Background Sources}: We also used the NASA/IPAC Extragalactic
Database (NED\footnote{Available on the World Wide Web at
http://nedwww.ipac.caltech.edu.})  to search for background source
counterparts for the remaining eleven discrete X-ray sources
identified in our survey.  We adopted a radius of 6$\arcsec$ for this
search but no counterparts were identified for any of the eleven
sources. To estimate the number of detected background sources that
are seen in projection beyond NGC 7793, we use the relation given by
\citet{Campana01} for the number $N$ of background sources greater
than a flux density $S$ per square degree, which may be expressed (in
CGS units) as

\begin{equation}
\mbox{$N$ ( $>$ $S$)} = 360 \left( \frac{S}{2 \times 10^{-15}} \right)^{-0.68}. 
\label{BackEqn}
\end{equation}

If we consider the entire ACIS-S3 chip (with a field of view of
8.3$\arcmin$ $\times$ 8.3$\arcmin$) and assume a limiting absorbed
flux of 1.15$\times$10$^{-15}$ ergs cm$^{-2}$ s$^{-1}$ for our
observation, we estimate that approximately ten background sources lie
within those bounds.  NGC 7793 covers approximately half of the
ACIS-S3 chip, so we adopt a contamination of $\sim$5 background
objects within the optical extent of the galaxy.

\par

{\it Sources Previously Blended and Now De-Blended by Chandra}: Finally, we 
describe our search for
sources which may have been blended by previous observations but have
been resolved with {\it Chandra}. Besides the resolved emission
components from the candidate microquasar N7793-S26, we find two such instances: the
first includes the two X-ray sources found within the error circle of
the source P13, the ULX, discussed previously.  The second instance
involves the sources CXOU J235802.8$-$323614 and CXOU
J235803.54-323643 which are located within 20$\arcsec$ and
11$\arcsec$, respectively, of the position of the source P11
identified by RP99.  The latter source is only slightly more luminous
than the former and it appears that the combination of emission from
both were identified as P11 in the PSPC observation.

\subsection{Quantile Analysis of the Spectral Properties of the Discrete 
X-ray Sources\label{QuantileSection}}

We adopted the quantile approach to a color-color diagram
\citep{Hong04}.  The quantile method determines the energy below which
a fixed percentage of events fall: colors are therefore determined from the 
ratios or differences of the resulting energies. Briefly, $E$$_X$ is defined as
the energy below which the net counts are X$\%$ of the total net counts; 
$E$$_{25}$, $E$$_{50}$ and $E$$_{75}$ then correspond to the energies below
which the counts are 25$\%$, 50$\%$ and 75$\%$, respectively, of the total counts.
Further, the quantile $Q$$_X$ is defined as 

\begin{equation}
Q_X = \frac {E_X - E_{low}}{E_{high} - E_{low}}
\end{equation}

\noindent
where $E$$_{low}$ and $E$$_{high}$ are the lower and upper boundary
energies respectively of the full energy band considered (in the present paper,
we have assumed values of $E$$_{low}$ = 0.2 keV and $E$$_{high}$ = 10.0 keV).
The grid is defined by 3*($Q$$_{25}$/$Q$$_{75}$) versus 
log ($Q$$_{50}$/(1-$Q$$_{50}$)) to separate the data as much as possible. The
appearance of the interpretative grid in model coordinates (e.g., $N$$_H$ versus 
$kT$, $N$$_H$ versus power law index) has a quashed appearance, reflecting the
spectral energy information truly available from the instrument.
The reader is referred to \citet{Hong04} for more information
about the quantile approach.

\par
\begin{figure}
\includegraphics[angle=-90,scale=0.70]{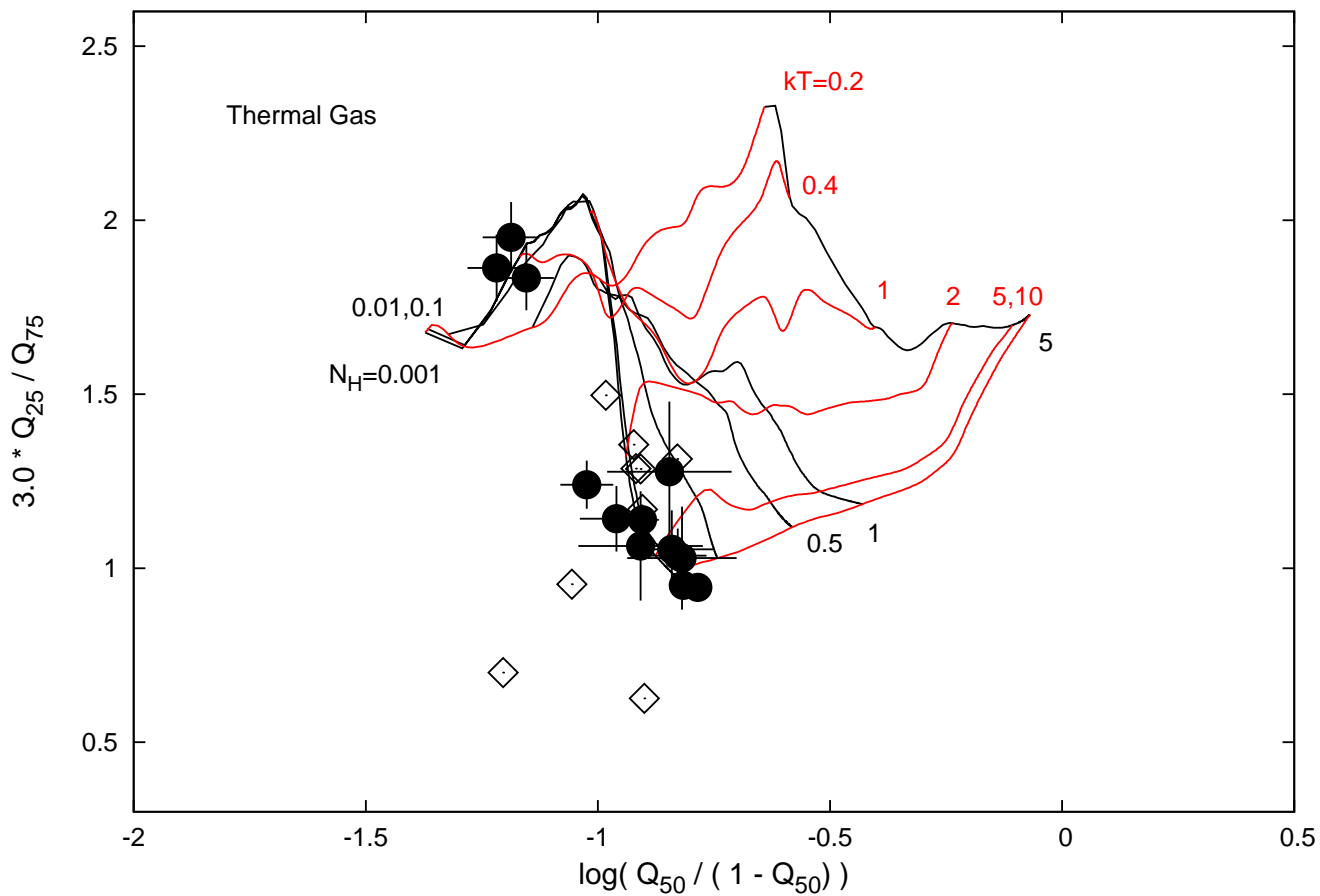}
\caption{Quantile plot for the detected sources in NGC 7793 using an 
optically-thin gas (APEC) model.  Note the two groups of points: a scatter of points at 
3*Q$_{25}$/Q$_{75}$$\sim$1.25 and  
three points at values of 3*Q$_{25}$/Q$_{75}$$\sim$2. The sources plotted with filled circles
exhibit 1$\sigma$ error bars. The sources plotted with open diamonds are weak
sources and the resulting errors are large, reducing the legibility of the grid. Error bars
are suppressed when the length of the bar covers at least one-half of the grid width. The units
for $N$$_H$ are 10$^{22}$ cm$^{-2}$ and for $kT$ are keV. See Section
\ref{QuantileSection} for a description of the model, the groupings of the points and further details.\label{fig9}}
\end{figure}
\begin{figure}
\includegraphics[angle=-90,scale=0.70]{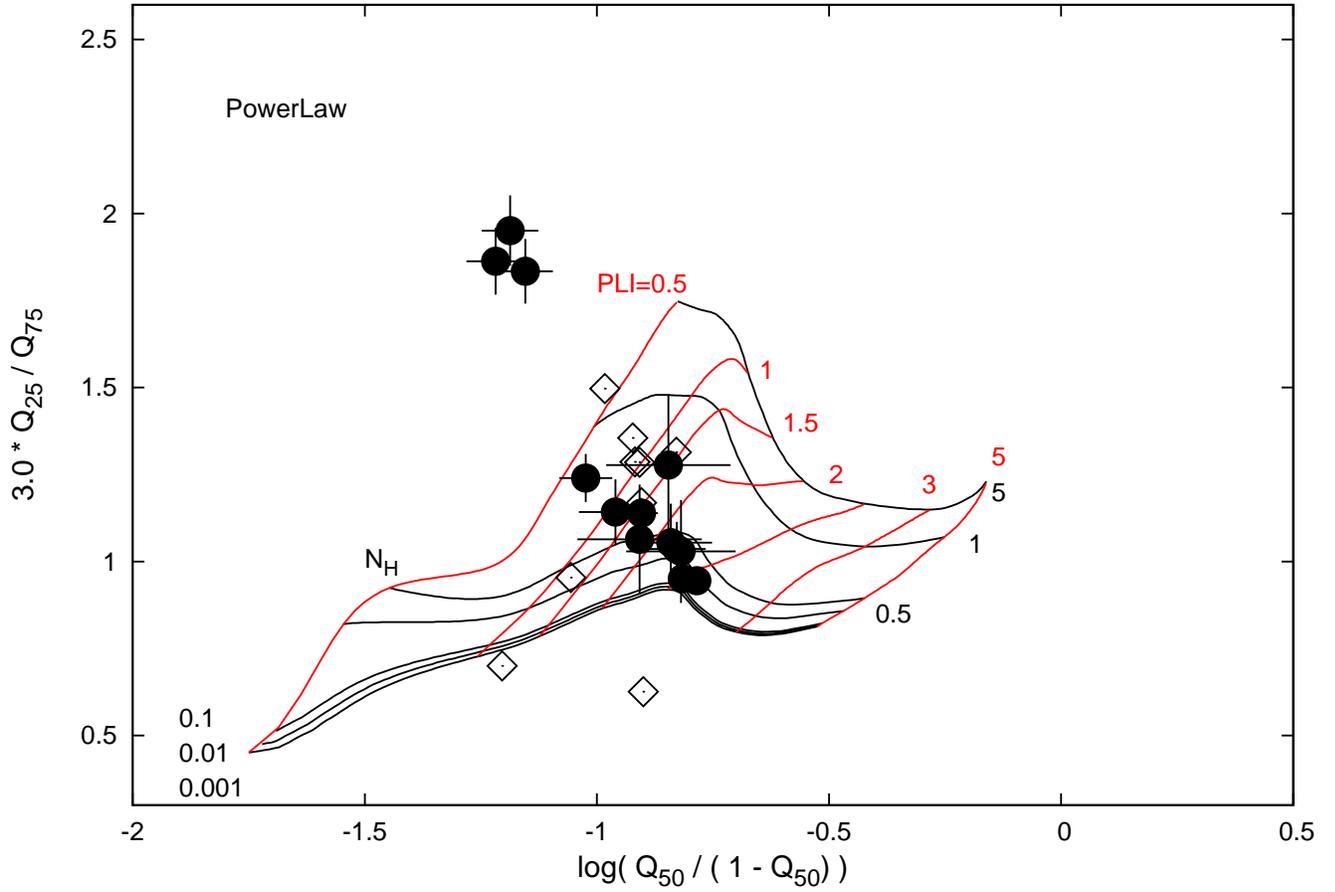}
\caption{Quantile plot for the detected sources
in NGC 7793 using a power law model. The plot symbols are defined in the caption
of Figure \ref{fig9}: the units for $N$$_H$ are 10$^{22}$ cm$^{-2}$ and the power law 
index ``PLI" is defined as $E$$^{-PLI}$.
See Section \ref{QuantileSection} and Figure \ref{fig9} for a description of the model,
the groupings of points and further details.\label{fig10}}
\end{figure}

Calculated values for $Q$$_{25}$, $Q$$_{50}$, $Q$$_{75}$, 
and 3*($Q$$_{25}$/$Q$$_{75}$) are presented in
Table \ref{Table5}. For example, in the case of the first tabulated source
CXOU J235743.8$-$323633, we have measured a value for $E$$_{25}$ = 1.14 keV
and the corresponding value for $Q$$_{25}$ is 0.096. 
In Figures \ref{fig9} and \ref{fig10} we present the quantile grids for the optically-thin
gas model APEC and power law models, respectively.  For both grids,
moving vertically in the grid crosses lines of equal N$_{\rm H}$ with
values of 0.001 (bottom), 0.01, 0.05,  0.1, 0.5, 1.0 and
5.0$\times$10$^{22}$ cm$^{-2}$ (top).  The APEC temperature $kT$
increments horizontally with values of 0.2 (left), 0.4, 1.0, 2.0, 5.0,
and 10.0 keV (right).  The power law photon indices include grid
values of 0.5 (left), 1.0, 1.5, 2.0, 2.5, 3.0, and 5.0 (right). 

\par

We have assumed a foreground Galactic column density toward NGC 7793
of $N$$_{\rm H}$ = 1.15$\times$10$^{20}$ cm$^{-2}$: this value
corresponds to a line of constant 3*($Q$$_{25}$/$Q$$_{75}$) of $\sim$1.  
If we momentarily ignore the error bars,
the plots show approximately two distinct groups: three points located 
near (log $Q$$_{50}$/(1-$Q$$_{50}$), 3*($Q$$_{25}$/$Q$$_{75}$))
$\sim$ (--1.2, 1.8) (which we shall denote as Group (i)) and a group located 
at $\sim$(--0.9, 1.0) (which we shall denote as Group (ii)). Considering Group (ii) 
first, we note that this group contains a mix of sources: the
counterpart to the HII region D2 (CXOU J235743.7$-$323633), the X-ray
counterpart to the candidate radio SNR R3 (CXOU J235746.7$-$323607), the
probable ULX (CXOU J234750.8$-$323726), two foreground stars (CXOU 
J235748.6$-$323234 and CXOU J235752.7$-$323309, which are associated with 
USNO 0574-1250312 and USNO 0574$-$1250339, respectively), and potentially about five 
background AGNs. The Group (ii) sources are broadly consistent with the column
density toward NGC 7793: sources in this group are also consistent with hard emission 
as indicated by their positions in both quantile plots.

\par

The three sources that belong to Group (i) are CXOU J235747.2$-$323523 
(the X-ray counterpart to the SNR N7793-S11 as discussed in Section 
\ref{OtherXraySourcesSection}), CXOU J235800.1$-$323325 and CXOU 
J235800.3$-$323455. While the identification of the X-ray counterpart to N7793-S11
seems clear, there are several possible physical interpretations possible for other the 
sources that belong in Group (i).
If an H~II region contains an SNR, then the
separation of the Group (i) points from the H~II regions in Group (ii)
may result from the strong line emission present in the spectrum of the SNR; the
excess absorption could be explained as circumstellar matter or
recombining nucleosynthesized matter.  Alternatively, if the H~II
region contains the remains of a massive star, the emission would be expected to
be much harder if dominated by an XRB and hence the point would appear
lower in the plot (that is, at higher ${\it kT}$).  We suspect that the Group (i) points
are mainly SNRs or other emission-line sources: alternatively, these sources may
also be XRBs illuminating adjacent clouds of interstellar material in NGC 7793.
We note that strong emission line sources are more difficult to interpret correctly
using the quantile approach because strong, low-energy line emission (at CCD
spectral resolution) can mimic a source with a continuous spectrum and {\it lower}
column density.

\subsection{Time Variability\label{VariabilitySection}}

To investigate time variability from the discrete X-ray source
population of NGC 7793, we examined two approaches: flux differences between our
{\it Chandra} observation and the {\it ROSAT} PSPC observation of RP99 as well as
variability within the {\it Chandra} observation itself.
For the differences in flux between the two observation epochs, we must consider 
the flux differences between
sources detected by both RP99 and the present work, sources detected
by RP99 but not detected by the present work, and sources detected by
the present work but not by RP99.  We adopted a common energy range
(0.2-2.4 keV) to make the flux comparisons as well as a commmon
spectral model (that is, an absorbed power law).
\par
We note that we have recovered all of the RP99 sources.  We have found
only one {\it Chandra} source that should have been detected by RP99
if it were active during the {\it ROSAT} PSPC observation, namely CXOU
J235806.6$-$323757 (see Section \ref{LuminousXraySourcesSection}).
We considered the seven discrete X-ray sources detected both in
this paper and by RP99.  Table \ref{Table6} presents estimates for the
{\it ROSAT} and {\it Chandra} luminosities of these sources using a
power law model with a photon index $\Gamma$=1.5 and a column density
of $N$$_H$=1.15$\times$10$^{20}$ cm$^{-2}$.  For CXOU
J235746.7$-$323607 we also included a luminosity estimate based upon
the {\it Einstein} observation of this source as described by
\citet{Harris94} and designated in that work as 2E 2355.2-3253.

Precise comments for time variability are hampered by the lower
quality PSF of {\it ROSAT} relative to {\it Chandra}.  The broader PSF
of {\it ROSAT} mixed diffuse emission into the discrete emission,
raising the overall count rate as well as altering the nature of the
spectrum, depending upon the size of the counts aperture used.  The
presence of the mixing may be noted because all of our {\it Chandra}
luminosity estimates are at least 10\% lower than the {\it ROSAT}
values.

Regardless, several sources stand out and merit a brief discussion.
CXOU J235806.6-323757 should have been detected, implying at least an
increase in flux by a factor of $\sim$30-50.  CXOU 235748.6-323234
decreased by a factor of $\sim$20 while CXOU J235800.1-323325
decreased by a factor of $\sim$6-7.  We therefore identify a total of three
variable discrete X-ray sources in NGC 7793: the remaining sources decreased
by modest amounts (factors of $\sim$2) or are constant within the
errors.

\par

To test for variability within the {\it Chandra} observation, we used
the standard {\it CIAO} tools to first barycenter the data and then extract light curves
for each detected point source. The light curves were binned into 60-second intervals
and these binned light curves were then run 
through a Bayesian variability detector (the CIAO tool 
{\tt glvary}). None of the sources exhibited statistically significant variability. An
examination of each light curve verified that the light curves of all of the discrete
sources were constant within the errors.

\subsection{Point Source Spatial Asymmetry\label{AsymmetrySubSection}}

We noted above that a majority of the point sources are located in the
eastern half of the galaxy and none are located in the northwestern
quadrant.  Such a peculiar distribution of sources is difficult to
reconcile with the general orderly and symmetric optical appearance of
the galaxy. Explanations for such a distribution include an excess
absorption toward the northwestern quadrant of the galaxy, a random probability
in the distribution of the sources, a 
gravitational interaction between NGC 7793 and another galaxy, 
dramatically lower effective exposure on this portion of the chip
during the observation and 
``patchy" star formation activity in NGC 7793. We rule out an excess absorption
toward the northwest quadrant based on inspection of maps of dust
column density toward NGC 7793 as provided by
\citet{Schlegel98}. These maps indicate a dust differential measure of
only 2\% across the face of NGC 7793, which is insufficient to
account for the observed asymmetry.  We consider the other four
explanations here in turn.

\par

We first consider a strictly random interpretation, that is, if most of
the sources are X-ray transients, by chance we observed NGC 7793 at an epoch
where sources in the northwestern portion of the galaxy were in an off
state.  Consider, for example, treating NGC 7793 as a dartboard.
Noting that approximately one-quarter of the galaxy is devoid of point sources,
we can calculate the probability (based on areas) that 20 point sources 
land outside one quadrant to be $\sim$(3/4)$^{20}$ $\approx$ 3$\times$10$^{-3}$.
If we instead consider on-and-off variability and assume that approximately eight
sources in one quadrant must be ``off" \footnote{The estimate of eight sources is based on 
the fact that because 20 sources are detected in approximately three-quarters of
the area of the galaxy, we may thus expect (4/3)$\times$20 $\approx$ 28 sources in
the whole galaxy.}, the probability of not detecting any source at all in this quadrant
is $\sim$(1/2)$^8$ $\approx$ 4$\times$10$^{-3}$. We readily acknowledge the 
{\it a posteriori} nature of our argument but we nonetheless consider the orders of magnitude
instructive. Having stated our interpretation, the difficulty with this
interpretation is the lack of sources in the northwest quadrant in all of the prior
{\it Einstein} and {\it ROSAT} observations as well as the {\it Chandra}
observation considered here. This result makes an explanation based on
short-term time-variability of the discrete X-ray sources less likely.

\par

A gravitational interaction with another member (or multiple galaxies)
of the Sculptor Group is another possibility for explaining the asymmetry: such an interaction
should also trigger enhanced massive star formation within NGC 7793.
The signposts of an elevated star formation rate have been revealed by
numerous observations of NGC 7793 including copious amounts of diffuse
radio continuum emission and diffuse [S II] emission from the galaxy's
disk \citep{Harnett86, Blair97}, the elevated infrared and blue
luminosities of this galaxy for its Hubble type \citep{Read99} and the
considerable population of resident OB associations and HII regions
\citep{Davoust80,Ferguson96}. 

\par

We have inspected tabulated information and distribution maps of the
Sculptor Group member galaxies as provided by \citet{Puche88} and
\citet{Karachentsev03} to identify a galaxy (or galaxies) which may be
interacting with NGC 7793.  The galaxy closest to the southeastern
edge of NGC 7793 is the Sculptor Diffuse Irregular Galaxy (SDIG, also
known as ESO349-G031) \citep{Laustsen77, Heisler97}: the HI mass of
the SDIG has been estimated to be 1.1$\times$10$^7$ M$_{\odot}$
\citep{Cesarsky77} which is more than an order of magnitude less than
the HI mass of NGC 7793. Assuming a distance of 4.1 Mpc to the SDIG
\citep{Karachentsev03}, the projected separation between this galaxy
and NGC 7793 is only 0.3 Mpc. Supporting evidence for a physical 
association between NGC 7793
and the SDIG is presented by \citet{Karachentsev03}, who argue that
the SDIG and a second dwarf galaxy (UA 442) are companions to NGC 7793
based on similar measured radial velocities for all three
galaxies. NGC 55, the nearest major Sculptor Group spiral galaxy to
NGC 7793, does not appear to be interacting significantly with NGC
7793.  While the HI mass of NGC 55 \citep[9$\times$10$^{8}$
M$_{\odot}$,][]{Puche91} more closely matches the HI mass of NGC 7793,
the projected distance from NGC 7793 is rather large (2.1 Mpc) if a
distance of 1.8 Mpc to NGC 55 is assumed \citep{Karachentsev03}. In
fact, \citet{Karachentsev03} argue that NGC 55 is instead associated
with a third major Sculptor Group spiral galaxy NGC 300. This
conclusion was also reached by \citet{Pietrzynski06}, who have
separately derived a distance of 1.9 Mpc to NGC 55. We can quantify the
likelihood of an interaction between NGC 7793 and either the SDIG or
NGC 55 by calculating the tidal index $\Theta$ \citep{Karachentsev99},
which may be defined as follows. If $M$ is the mass of the galaxy which is
suspected of interacting with a galaxy of interest and $D$ is the three-dimensional
separation between that galaxy and the galaxy of interest, then
\begin{equation}
\Theta  = \mbox{log} (M/D^3) + C, 
\end{equation}
where $C$ is a constant equal to $-$11.75 when $M$ is expressed in 
units of solar masses and $D$ is expressed in units of megaparsecs. If
$\Theta$ is calculated to be less than zero, then it may be safely concluded 
that the galaxy of interest and the suspected interacting galaxy are not
in fact interacting to a significant extent. We calculate values for $\Theta$ =
-3.14 and $\Theta$ = -3.76 in the cases of NGC 55 interacting with NGC 7793
and the SDIG interacting with NGC 7793, respectively. Based on these strongly 
negative values for the tidal index in both cases, we conclude that
a gravitational interaction is an unlikely explanation for the observed asymmetry.
Separately, we have also inspected GALEX data for NGC 7793 to search for any obvious
asymmetry in the ultraviolet morphology of the galaxy (as might be expected by
lopsided star formation) but we find no evidence for an asymmetric appearance at
that wavelength.

\par 

\begin{figure}
\includegraphics[scale=0.85]{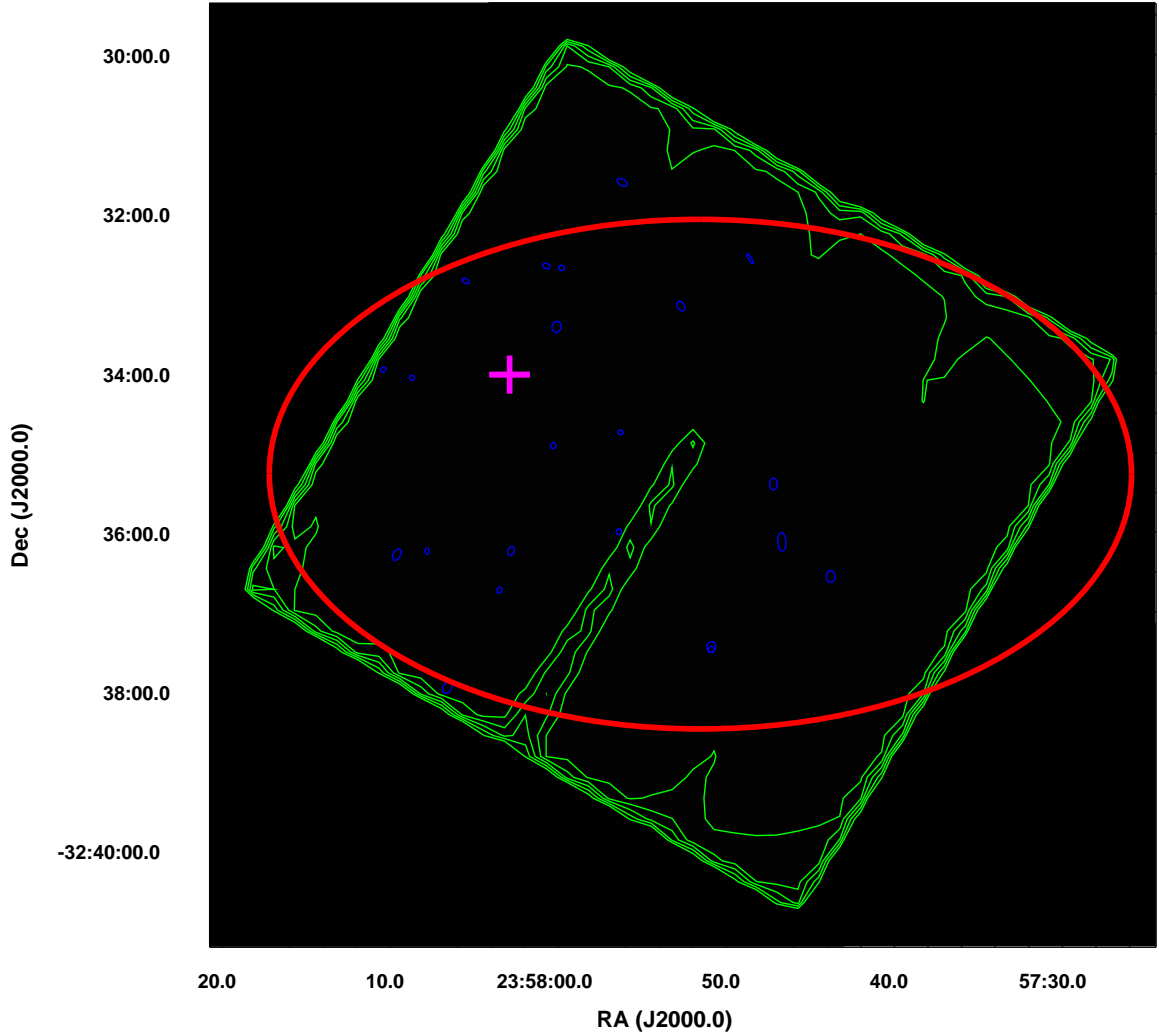}
\caption{Contour plot showing levels of exposure for the ACIS-S chip during the
observation of NGC 7793. The green contours depict levels corresponding to
90\%, 85\%, 80\%, 75\% and 70\% (respectively) of the peak exposure value of 
$\sim$3$\times$10$^7$ cm$^{-2}$ s counts photon$^{-1}$. Similar to Figure \ref{fig1},
the detected sources are indicated with blue ellipses, the red ellipse indicates the optical 
extent of NGC 7793 and the magenta cross indicates the aimpoint of the observation. We
argue that the lack of detections of discrete X-ray sources in the northwestern quadrant
of NGC 7793 cannot be explained by simply a significantly lower effective exposure
in that portion of the ACIS-S3 chip. See Section \ref{AsymmetrySubSection}.
 \label{fig11}}
\end{figure}

Next, we considered the possibility that the effective exposure time of the
portion of the ACIS-S3 chip that sampled the northwestern quadrant of NGC 7793
was significantly lower than for the rest of the chip, thereby leading to an observed deficit
of discrete X-ray sources in this part of the galaxy.  In Figure \ref{fig11}, we present
a contour plot showing the effective exposure for the ACIS-S3 chip during the observation:
for illustrative purposes, we also include the positions of the detected discrete X-ray sources,
the location of the aimpoint for the observation and finally an ellipse that spans the optical
extent of the galaxy. We argue that in fact the effective exposure time for this portion of
the chip is not significantly lower than for the rest of the chip and that therefore a different
effective exposure time cannot account for the observed asymmetry of the detected
discrete X-ray sources.

\par

Lastly, regarding the ``patchy" star formation scenario, we have also considered the work of 
 \citet{Smith84}, who described how the star formation
activity in NGC 7793 is stochastic and occurs only in large irregular
``patches"; such ``patchy" activity may also explain the observed
asymmetric distribution of sources.  \citet{Smith84} accurately modeled the star formation
activity in this galaxy using a stochastic self-propagating star
formation model without an imposed spiral modulation. Those authors
also commented that ``flocculent" galaxies such as NGC 7793
\citep[as described by][]{Elmegreen81} in general lack spiral modulation to star
formation and feature a ``patchy" arm structure. The lack of spiral
modulation in NGC 7793 is also supported by the absence of prominent
emission from the galaxy's nucleus at any wavelength.  If the star
formation in the northwest quadrant happened to belong to one large
``patch,'' then the lack of X-ray sources not only at the {\it
Chandra} epoch but also at the epochs of the prior X-ray observations
can be explained.  However, this scenario lacks the full status of an
explanation given that we do not have a method to date such a
``patch."

\par

Therefore, at the present time we cannot provide a clear explanation
for the observed asymmetry of discrete X-ray sources in NGC 7793. Random
probability seems more likely than such 
explanations such as a gravitational interaction with another galaxy or stochastic
star formation but additional study and analysis is required.

\section{Diffuse Emission\label{DiffuseSection}}

 A spectrum of the diffuse emission was extracted following the method
described by \cite{Schlegel03B}.  Point sources were removed by
screening out all events within a radius that enclosed $>$95\% of the
PSF at the detected position of each point source.  The screening
radius was increased to match the increase in the PSF with off-axis
angle.  The resulting holes were filled in by randomly selecting from
an annulus surrounding each source the approximate number of events
that would have been present based on the count rate in the annulus.\footnote{We 
could ignore the regions surrounding point sources but our
analysis of the diffuse emission is part of an investigation of the spatial
distribution of the diffuse emission in face-on spirals for which we do not
want holes (Schlegel et al. 2011, in preparation).}
This process assumes only spectral and spatial uniformity of the
diffuse emission on spatial scales of $\sim$20-30$''$.  The inner
radius of the annulus used for the back-fill was twice the outer
diameter of the source screening radius to reduce the probability that
the selected events were point source events scattered by the wings of
the PSF.  The annulus was 20$''$ wide.  Annulus overlaps with other
point sources were minimal, but were avoided by sampling events from
the non-overlapping portions of the annulus.  A radial profile of the
diffuse emission was obtained from azimuthal sums in 10$''$-wide
annuli centered on the nucleus.  The diffuse spectrum was then
extracted using an aperture with a radius of $\sim$3$'$.7 determined
by the point at which the diffuse profile joined the local
(non-galaxy) background.  An estimate of the impact of the backfill
may be determined by summing the extraction areas of the point sources
and dividing by the extraction area of the entire galaxy.  For NGC
7793, those values are 1900 arcsec$^2$ and 1.55${\times}$10$^5$
arcsec$^2$, respectively, for an impact ratio of $\sim$1.2\%.

\par

The spectral fit of the diffuse emission spectrum was carried out in
two steps.  First, the background was fit using a variety of continuum
and line features to achieve a good fit as described in
\citet{Schlegel03B}.  Second, the diffuse spectrum + background was
fit simultaneously with the background by adopting for the background
the best-fit parameters determined in the first step\footnote{This
approach preserves the proper statistical distribution in contrast to
a background-subtracted spectrum and has become common practice in the
field since \cite{Cash79}.}.  A version of the optically-thin thermal plasma model
APEC \citep{Smith01} which is known as ``VAPEC" and which includes variable
elemental abundances as determined using recent atomic physics was used to 
fit the spectrum.  Abundances were allowed to
vary but if an abundance was found to have an error that included
unity, the abundance value was reset and fixed at unity. The known
absorbing column density toward NGC 7793 was adopted and the corresponding
fit parameter was fixed at that value.  The background spectrum contained a large
fluorescent Si feature at $\sim$1.78 keV.  The background fit easily
matched the data, but the result slightly over-corrected the source
spectrum because of photon statistics.  We excised the data at this
location: this excision has a negligible effect on the fit as the
dominant emission from the hot gas occurs in the 0.5-1 keV band.  We
also included a power law component to account for any hard emission
present from unresolved point sources. The results of our fits are summarized
in Table \ref{diffuse-fit}.
\par
\begin{figure}
\includegraphics[angle=-90,scale=0.7]{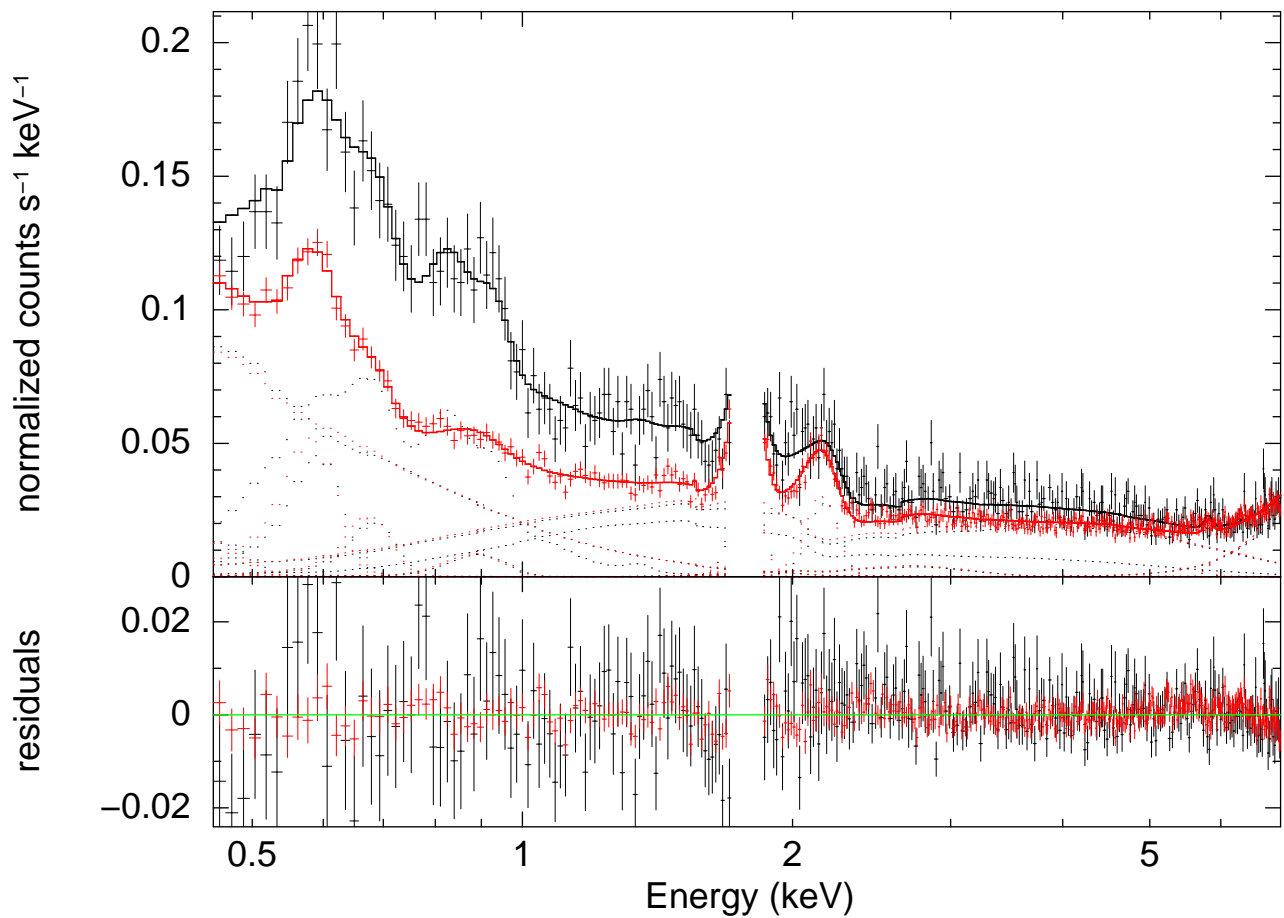}
\caption{Extracted spectrum of the diffuse X-ray emission of NGC 7793.
The source+background spectrum is shown in black while the background
spectrum is shown in red. 
Dotted lines represent the individual model components.
The spectra have been fit using the VAPEC model: the column density
has been fixed at the value corresponding to the Galactic column
density toward NGC 7793 ($N$$_H$ = 1.15$\times$10$^{20}$ cm$^{-2}$).
The background spectrum contained a large fluorescent feature at 1.78
keV: data was excised at this location. See Section 
\ref{DiffuseSection}.}
\label{fig12}
\end{figure}

In Figure \ref{fig12} we present the extracted spectrum of the diffuse emission 
as fit with the VAPEC model. The fitted temperature of the diffuse spectrum of 
NGC 7793 was found to be ${\it kT}$ = 0.19$^{+0.03}_{-0.02}$ keV and 
$N$$_{\rm H}$ = 3.6${\times}$10$^{21}$ cm$^{-2}$, ${\it kT}$ = 0.22${\pm}$0.02 keV
for the dual APEC fit, or ${\it kT}$ = 0.25${\pm}$0.02 keV if $N$$_{\rm H}$ is fixed 
at the Galactic column density. These values are lower than the temperatures 
derived by RP99 (namely ${\it kT}$ $\sim$0.8-1.1 keV) using typical thermal models 
such as thermal bremsstrahlung and the Raymond-Smith thermal plasma. 
The most likely explanation for the discrepancies in the fitted temperature is 
the mixing of the diffuse emission with point source emission from the broad
wings of the {\it ROSAT} PSPC PSF.  For the PSPC, a circle enclosing
90\% of the flux was 0.9$\arcmin$ in diameter at 1 keV but
2.7$\arcmin$ in diameter at 1.7 keV. RP99 adopted an aperture of
1$\arcmin$ so sources with harder spectra and hence larger PSFs would
preferentially contribute flux above a PSF radius of
$\sim$1.1$\arcmin$. In addition, weak unresolved point sources would
blend to form a brighter ``diffuse" component.
\par
\begin{figure}
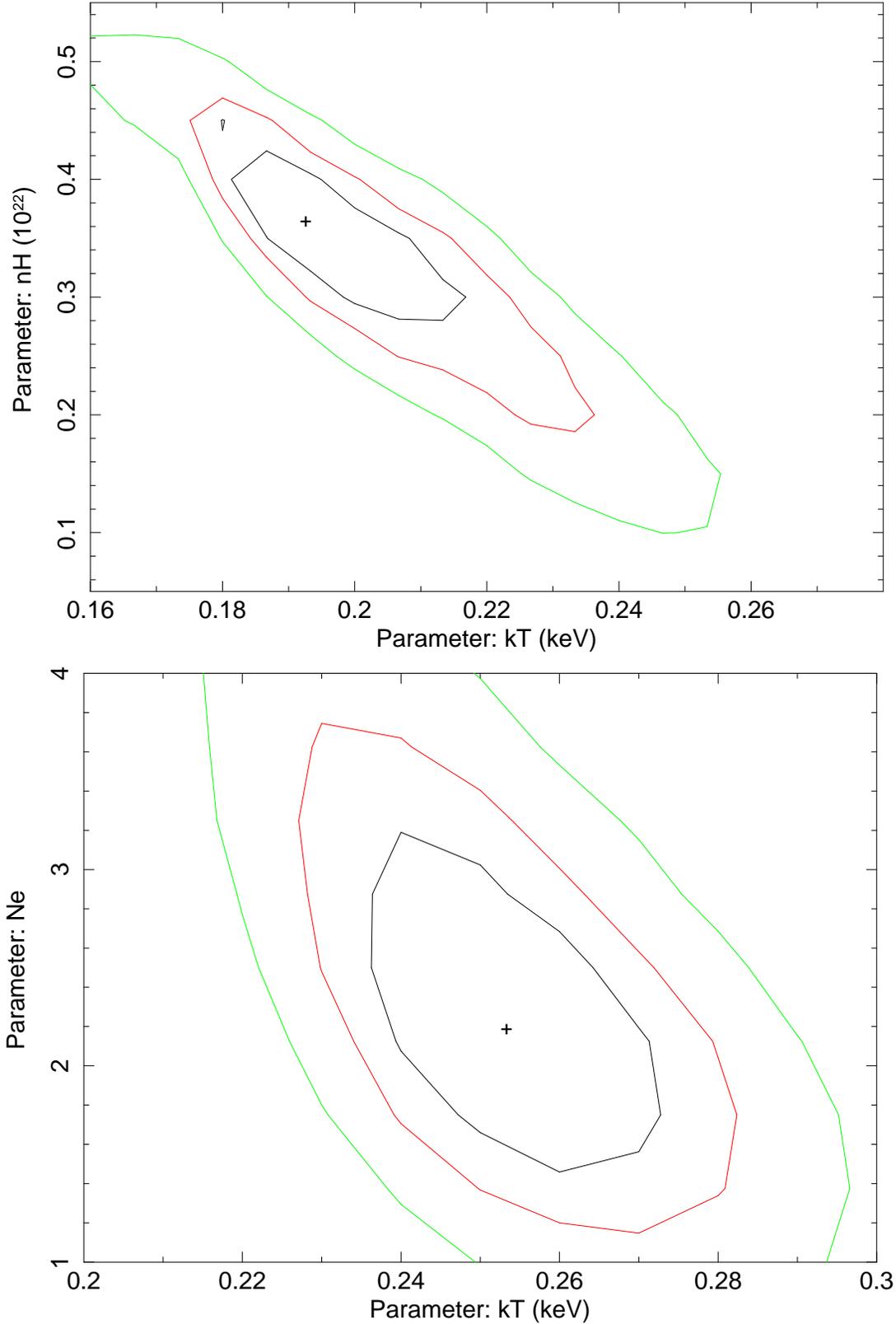

\epsscale{1.2}
\includegraphics[angle=-90,scale=0.6]{f13a.ps}
\includegraphics[angle=-90,scale=0.6]{f13b.ps}
\caption{\footnotesize{(top) Confidence contours for fit (as shown in Figure
\ref{fig12}) to the extracted spectrum of the diffuse emission; (bottom)
contours for the fixed $N$$_{\rm H}$ VAPEC model. The contours are
plotted at the same levels as in Figure \ref{fig3}. For neon, we
have plotted this parameter in units of solar abundance. See Section 
\ref{DiffuseSection}.}\label{fig13}}
\end{figure}

The fitted temperature of the diffuse emission is similar to the results from fits to the
diffuse components of other nearby galaxies: in Table~\ref{diffuse-table} 
we list several values measured for the diffuse emission from other galaxies
for comparison. NGC
7793 stands out for the absence of a second, hotter diffuse component.
We attempted to fit a second APEC model (i.e., VAPEC + VAPEC +
Power Law), but the model normalization of the second APEC component
was consistent with zero.  A second APEC component was non-zero only
if the $N$$_{\rm H}$ was fixed at the known Galactic column.  We expect
that a longer exposure would lead to sufficient statistics to separate
a hot component from the background emission without the necessity of
additional model constraints. The abundance of each element was permitted to 
vary, but only the
abundance of Ne in the fixed $N$$_{\rm H}$ VAPEC model was significantly
different from 1.0: in this case, the best-fit value for the abundance was
2.23$^{+0.74}_{-0.66}$. In Figure \ref{fig13} we present confidence contour
plots for the column density and the fitted temperature for the APEC fit (left) 
as well as the fitted temperature and neon abundance for the VAPEC fit in
which the column density was fixed at the known Galactic value (right). 
\par
We also modeled the spectrum as a sum of unresolved Gaussians plus a
power law with emission Gaussians at 0.60, 0.75, and 0.89 keV,
corresponding approximately to emission lines attributed to O~VII,
Fe~L shell, and Ne~IX.  The presence of these lines is expected in hot
diffuse gas.  The model fits (with both $N$$_{\rm H}$ free and fixed) were
consistent with the APEC+Power Law model, but applying Occam's razor,
this model is penalized by the extra components and constraints
necessary to obtain a good fit. While the multi-Gaussian fit implies the presence
of specific emission lines, the VAPEC model provides the statistically acceptable
fit and we consider it the best-fit model. 
\par
For the adopted Galactic column density and the VAPEC model, we
calculate absorbed and unabsorbed fluxes for the diffuse emission of
$\sim$5.4$\times$10$^{-13}$ and $\sim$5.5$\times$10$^{-13}$ erg
s$^{-1}$ cm$^{-2}$, respectively, in the 0.5-2 keV band. For the
assumed distance to NGC 7793, the unabsorbed flux corresponds to a
luminosity of $L$$_X$$\sim$3.3$\times$10$^{38}$ erg s$^{-1}$. 

\section{Luminosity Function of Discrete X-ray Sources\label{LFSection}} 

\begin{figure}
\epsscale{0.85}
\includegraphics[angle=-90,scale=0.7]{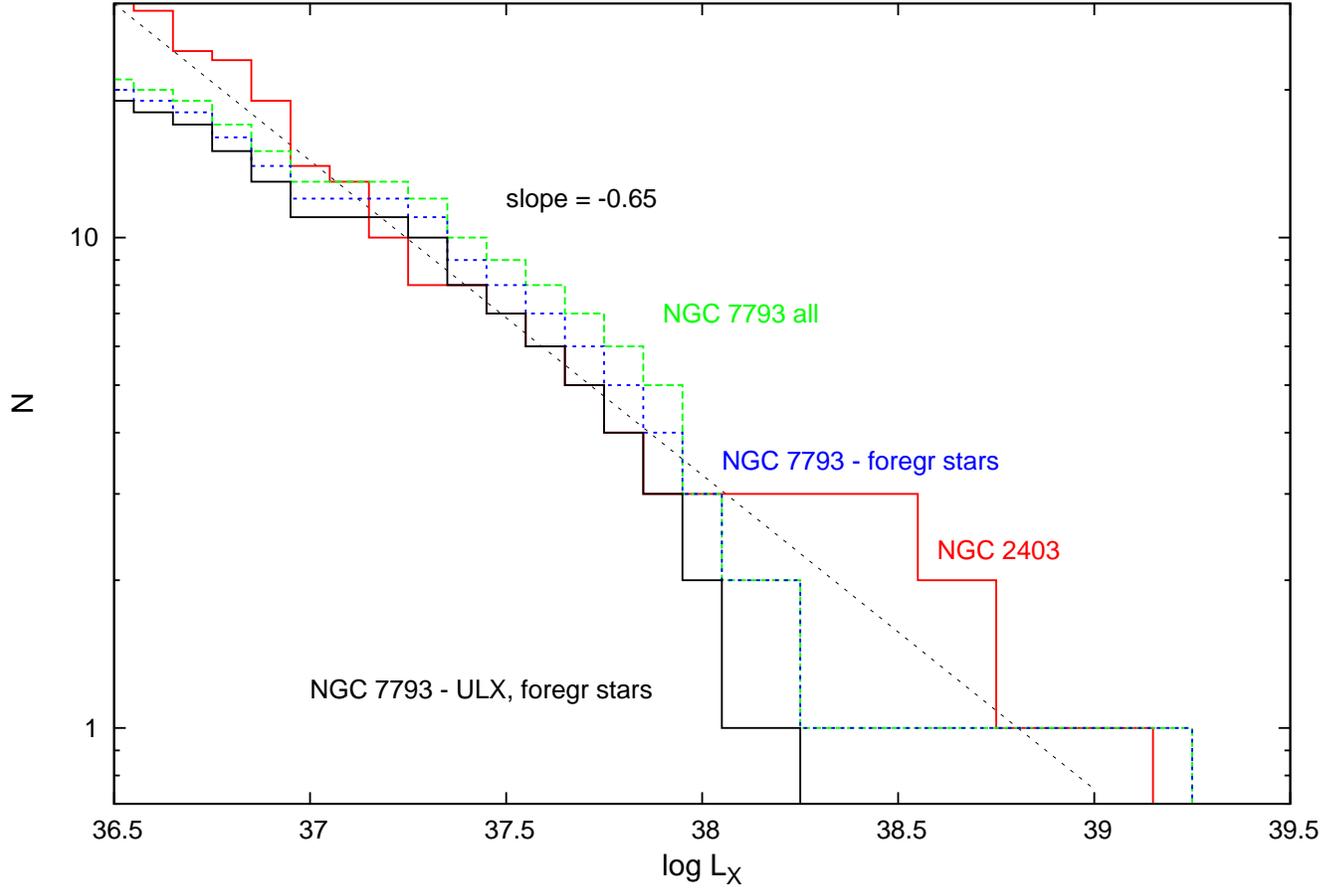}
\caption{Luminosity function for all discrete sources within the
optical extent of NGC 7793 (green dashed online), all sources except
the two known foreground stars (blue dashed online), and all sources
after excluding the foreground stars and the ULX (black line).  The
luminosity function of the discrete sources in the galaxy NGC 2403
\citep{Schlegel03A} is shown for comparison (red line online).  The
dashed line represents a slope of --0.65. See Section 
\ref{LFSection}.\label{fig14}}
\end{figure}

A plot of $N$ (defined as the number of sources with luminosities in excess
of the luminosity $L$$_X$) versus log $L$$_X$ for the NGC 7793 discrete sources (with
luminosity units of 10$^{38}$ erg s$^{-1}$) is shown in Figure
\ref{fig14}. Several functions are plotted, including the complete
luminosity function for NGC 7793, the luminosity function without the two known
foreground stars, and the function without the ULX and the two foreground
stars.  For comparison,
the luminosity function of NGC 2403 \citep{Schlegel03A} and a line of
slope $\Gamma$ = $-$0.65 are also shown.  The excluded ULX is very bright and its
inclusion leads to a long tail in the distribution, potentially
requiring a two-component fit.  We discuss the complete and the
ULX-less functions solely for comparison with the luminosity functions
of other galaxies.
\par
A linear fit to the complete function (log $N$ versus log $L$$_X$) yields a slope of
$\Gamma$ $\sim$$-$0.62$\pm$0.2, but the residuals are large (${\chi}^2/{\nu}
{\sim}$1.24).  For the ULX-less luminosities, the linear slope has an
absolute value of $\Gamma$ = $-$0.65$\pm$0.11 (${\chi}^2/{\nu} {\sim}$1.03).  Both
values are comparable to the log $N$-log $S$ slopes for NGC 6946 ($\Gamma$ $\sim$$-$0.64;
\citet{Holt03}) or NGC 2403 ($\Gamma$ $\sim$$-$0.59; \citet{Schlegel03A}) but are
dissimilar to the slopes of IC~5332 or M83 ($\Gamma$ = $-$1.30$\pm$0.31 and $\Gamma$ =
$-$1.38$\pm$0.28, respectively; \citet{Kilgard02}). 
\par
Note that the slope of $\Gamma$ = $-$0.65 is well-defined in the narrow luminosity
range of $\sim$37.1 $<$ log L$_{\rm X}$ $<$ $\sim$37.8 in contrast to
the slope of NGC~2403 (determined over the range $\sim$36.3 $<$ log $L$$_X$ 
$<$ $\sim$39.0 -- see \citet{Schlegel03A}).  There are a deficit of sources in NGC 7793 at
low ($<$37.0) and high ($\sim$38.3-39.0) log $L$$_{\rm X}$.  Both galaxies
occupied a similar area on the ACIS CCDs, so any loss of sources at low
L$_{\rm X}$ does not provide an explanation.  If one or two luminous
LMXBs were off during the observation epoch, the deficit at log
L$_{\rm X}$ $\sim$ 38.3 to $\sim$ 39.0 is readily explained.  At the present time, we have
information on the time variability of only the most luminous sources
(see Section \ref{VariabilitySection}).
\par
In previous work, differences between
the slopes of the luminosity functions of the different galaxies may
be reconciled by applying corrections for star formation rates. This
result was demonstrated by \citet{Grimm02} who aligned seemingly
discrepant luminosity functions after applying corrections for star
formation rates in these galaxies. In the case of NGC 7793, such a
correction is small \citep{Storchi94}. In Table \ref{lumfunction-table}, we present
estimates of both $\Gamma$ and star formation rates (in units of $M$$_{\odot}$ yr$^{-1}$)
for five galaxies, including NGC 7793. In contrast to the work of \citet{Grimm02}, 
we find no obvious correlation between the
values of the slopes and the star formation rates of the five galaxies we list in
Table \ref{lumfunction-table}.

\section{The Multi-Wavelength Properties of the SNR Population of 
NGC 7793\label{SNRSection}}

Lastly we comment on the multi-wavelength properties of the SNR population 
of NGC 7793. As noted previously, prior to the present work a total of 32 SNRs
had been identified in this galaxy by previous surveys conducted at X-ray, optical
and radio wavelengths \citep{Blair97,Read99,Pannuti02}. In Section 
\ref{DiscreteSection}, we presented a spectral analysis of CXOU 
J235746.7$-$323607, the X-ray counterpart to the candidate radio SNR
R3 that was identified by \citet{Pannuti02} and concluded that the X-ray
source was time-variable. We also exclude this source from our discussion here
of the multi-wavelength properties of the SNR population of NGC 7793, reducing
the size of the sample of sources to 31. 

By virtue of its superior angular resolution, {\it Chandra} potentially yields a 
significant improvement over {\it ROSAT} (which imaged NGC 7793 previously) in
studies of the SNR population of an external galaxy. To identify X-ray counterparts
to these known SNRs, we have cross-correlated the list of discrete X-ray sources
detected in this galaxy (see Table \ref{Table2}) with the positions of the known
SNRs (as described in Section \ref{OtherXraySourcesSection}). We have clearly
detected only one additional SNR, the optically-identified SNR N7793-S11, which
has also been detected in the radio by \citet{Pannuti02}. To investigate the
scenario where the X-ray counterparts to the SNRs may be extended and faint
which could be missed by {\tt wavdetect}, we extracted counts at the locations of
all of the 30 remaining SNRs. For the optically-identified SNRs we used apertures
that corresponded to the sizes of the angular extents of these SNRs (ranging in
size from $\sim$2$\arcsec$ to $\sim$10$\arcsec$ in radius) while for the candidate
radio SNRs we used apertures that were 3$\arcsec$ in radius. We did not detect
any additional SNRs to a limiting count rate of $\sim$1$\times$10$^{-4}$ cts s$^{-1}$:
we note that this search is further confused by diffuse X-ray emission from the disk
of NGC 7793 as well. To estimate a corresponding limiting luminosity for this
count rate, we consider the work of \citet{Long10} who conducted an X-ray survey of the
SNR population of the nearby face-on spiral galaxy M33 with {\it Chandra}, known
as the {\it Chandra} ACIS Survey of M33 (ChASeM33 -- see \citet{Plucinsky08}).
Those authors assumed a soft thermal ($\it{kT}$ = 0.6 keV) spectrum with a sub-solar
metal abundance of 0.5 for calculating the luminosities of detected X-ray counterparts
over the energy range of 0.35 - 2.0 keV. Assuming the same model, considering the
same energy range and adopting the nominal Galactic column density toward NGC
7793 of $N$$_H$ = 1.15 $\times$ 10$^{20}$ cm$^{-2}$, we calculate a limiting
luminosity for our search of $L$$_X$ $\sim$ 7.6$\times$10$^{35}$ ergs s$^{-1}$.
The survey conducted by \citet{Long10} identified 7 (out of 131 SNRs observed by the
survey) with X-ray luminosities in excess of this limit for a detection rate of 5\%:
this value closely matches our detection rate of one SNR detected (out of 31
SNRs observed) of 3\%.

\begin{figure}
\plotone{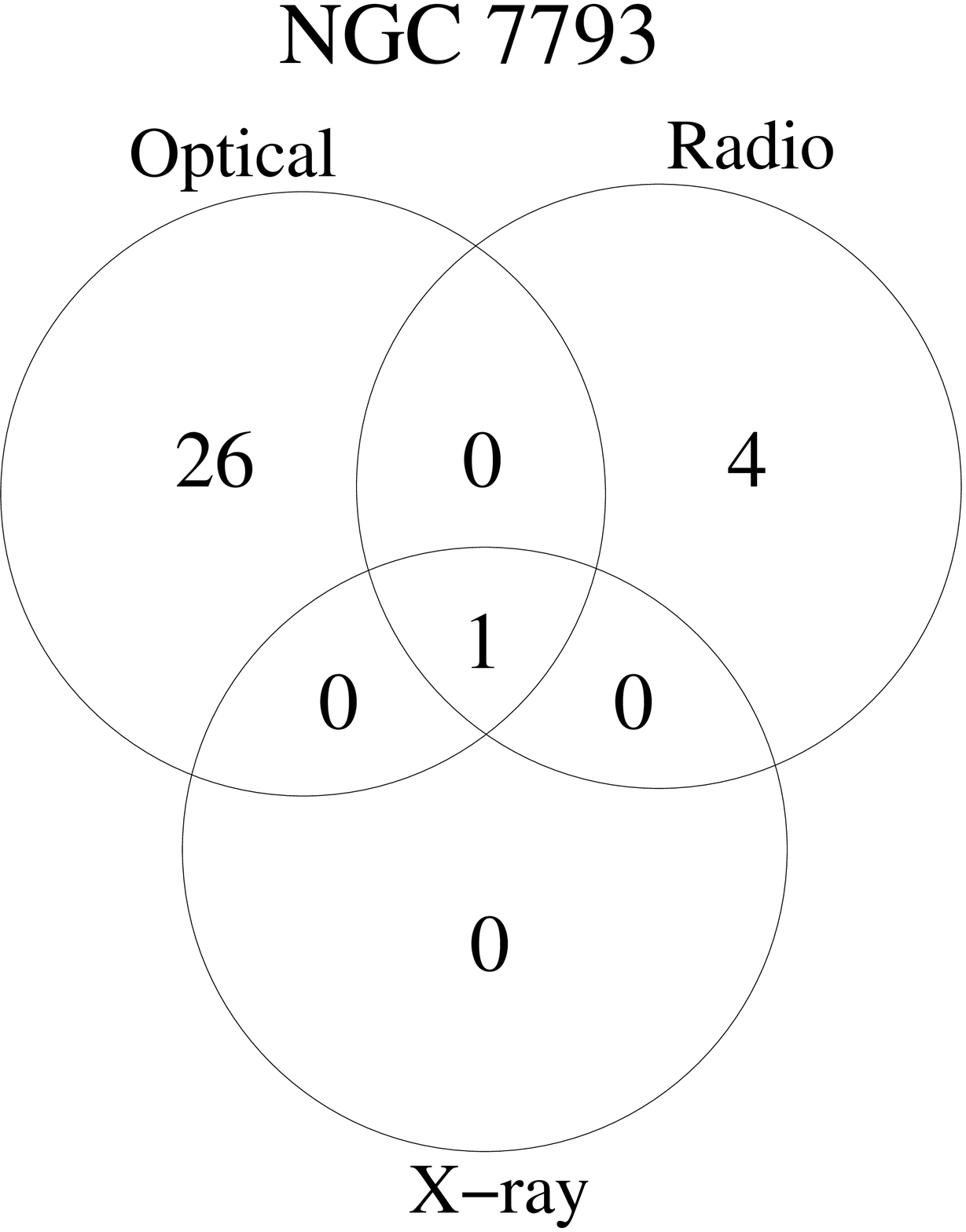}
\caption{Venn diagram showing the overlap of
the detections of SNRs in NGC 7793 at X-ray, optical and radio
wavelengths. See Section \ref{SNRSection}. \label{fig15}}
\end{figure}

We find therefore that we can reconcile the detection rate of X-ray counterparts to
SNRs in NGC 7793 with the detection rate of X-ray counterparts to SNRs in M33 as
presented by \citet{Long10}. For comparison purposes,
we note that over ten Galactic SNRs are known to have unabsorbed X-ray
luminosities that exceed the limiting luminosity attained by our {\it
Chandra} observation:\footnote{See ``The {\it Chandra} Supernova
Remnant Catalog" (http://hea-www.harvard.edu/ChandraSNR).} it is possible that
the discrepancy between NGC 7793 and the Milky Way may be simply due to the
lower mass and star formation rate of NGC 7793 compared to the Milky Way. Figure
\ref{fig15} presents a Venn diagram summarizing the overlap of
detections of the 31 SNRs identified at multiple wavelengths in NGC 7793 as updated with
the results of the present work. 

\citet{Pannuti07} presents and
discusses wavelength-dependent selection effects in our search for
SNRs in a sample of nearby galaxies using {\it Chandra} observations.
\citet{Long10} also discuss multi-wavelength properties of the sample of
known SNRs in M33: those authors describe the importance of local 
gas density in dictating the X-ray properties of SNRs as well as consider how optical
morphology and environment of the SNR may affect its detectability in the X-ray. 
The reader is referred to both of these works for a more complete discussion of
multi-wavelength properties of SNRs. We note that the detection of X-ray emission from
N7793-S11 is significant in that this source is one of only a very few
extragalactic SNRs located beyond the Local Group which have been
detected in the X-ray, optical and radio bands.

\section{Conclusions\label{ConclusionsSection}}

The conclusions of this paper may be summarized as follows:
\par

1) We detected 22 discrete X-ray sources within the optical extent of
NGC 7793 in a 49094 sec exposure.  The sources are significant at the
$\sim$3$\sigma$ level or greater and correspond to a limiting
unabsorbed luminosity of $\sim$3$\times$10$^{36}$ ergs
s$^{-1}$ over the 0.2-10.0 keV energy range.

\par

2) Four sources had a sufficient number of counts to allow spectral
fitting.  Acceptable fits were derived using either a power law, a
bremsstrahlung model, an APEC model or a disk blackbody model plus 
zero-width Gaussians to simulate unresolved lines.  Column densities were 
generally higher than the known Galactic value toward NGC 7793 -- in fact, 
fits using the known Galactic value were generally poor.  
Our derived fit to the extracted spectrum of the ULX
using the DiskBB model returns a value for $kT$$_{in}$ of
approximately 2.0 keV, consistent with the interpretation provided by
RP99 that this source is an XRB featuring a $\sim$ 10 $M$$_{\odot}$
mass black hole.  All sources were investigated using a quantile
color-color plot.  Time variability was investigated through
comparisons of the fluxes between our {\it Chandra} data and the RP99
results.  Three sources were shown to have varied by factors of
$\sim$6 to 30 or more.

\par

3) We searched for counterparts at multiple wavelengths for the
detected X-ray sources. Based on our search, we have identified
counterparts to one SNR, one HII region and two foreground stars;
the remaining sources are likely to be XRBs and luminous X-ray SNRs 
native to NGC 7793 and background galaxies seen through the disk of 
the galaxy.  The detected SNR -- N7793-S11 -- is also detected in the optical
and radio, making it one of the few SNRs located outside of the Local Group
to be detected at all three wavelengths. We have also ruled out
the possibility that the candidate radio SNR R3 is a single SNR.

\par 

4) A remarkable asymmetry is seen in the distribution of X-ray sources
in this galaxy, with the large majority seen in the eastern half. Possible
explanations for this asymmetry include a gravitational interaction with
a nearby galaxy or stochastic star formation. 

\par

5) The fitted temperature of the diffuse emission is
${\it kT}$=0.253$^{+0.018}_{-0.015}$ keV, lower than the temperature
measured by RP99. The discrepancy can be explained by the significant
mixing of the diffuse emission with point source emission from the
broad wings of the {\it ROSAT} PSPC PSF.

\par

6) We constructed the luminosity function for the detected discrete
X-ray sources in NGC 7793. If the known ULX in this galaxy is
excluded, the absolute value of the fitted slope is $\Gamma$=$-$0.65$\pm$0.11, but
the shape is linear over a small range. If the ULX is included, the absolute
value of the fitted slope becomes $\Gamma$=$-$0.62$\pm$0.2 but the residuals of
the fit are much larger.

\acknowledgments

We thank the referee for many helpful comments that have significantly
improved the quality of this paper.
We thank A.M.N. Ferguson for kindly sharing her H$\alpha$ and R-band
images of NGC 7793 for the purposes of this research.
T.G.P. acknowledges useful discussions with Nebojsa Duric about
properties of SNRs and Phil N. Appleton about the
peculiar distribution of the discrete X-ray sources in NGC 7793.  
T.G.P. also thanks Lauren Jones for reviewing this manuscript and
commenting on the phenomenon of star formation in galaxies. This
research has made use of NASA's Astrophysics Data System as well as
the NASA/IPAC Extragalactic Database (NED) which is operated by the
Jet Propulsion Laboratory, California Institute of Technology, under
contract with the National Aeronautics and Space Administration. This
research has also made use of data obtained from the High Energy
Astrophysics Science Archive Research Center (HEASARC), provided by
NASA's Goddard Space Flight Center. This work is supported by Chandra
Grant GO3-4104Z and this research has made use of software provided by the
{\it Chandra X-ray Center (CXC)} in the application package {\it CIAO.}

Facilities: \facility{CXO(ACIS)}.

\clearpage

\clearpage

\clearpage
\begin{deluxetable}{lcc}
\tabletypesize{\scriptsize}
\tablewidth{0pt}
\tablecaption{Summary of Gross Properties and {\it Chandra} Observation of NGC 7793\label{Table1}}
\tablehead{
\colhead{Property} & & \colhead{Reference} 
}
\startdata
Nucleus \\
~~RA (J2000.0) & 23$^h$ 57$^m$ 49$^s$.83  & (1) \\
~~Dec (J2000.0) & -32$^{\circ}$ 35$'$ 27$''$.7 & (1) \\
~~Galactic Longitude $l$ ($^{\circ}$) & 4.52 & (1) \\
~~Galactic Latitude $b$ ($^{\circ}$) & $-$77.17 & (1) \\
Galaxy \\
~~Observed Diameter $D$$_{25}$ (arcmin) & 9.2 & (2) \\
~~Axial Ratio $d$/$D$ & 0.70 & (2) \\
~~Hubble Type & SA(s)d HII & (1) \\
~~HI Mass (M$_{\rm{\odot}}$) & 9.1$\times$10$^8$& (3) \\
~~Distance (Mpc) & 3.91 & (4) \\
~~Inclination $i$ ($^{\circ}$) & 50 & (2) \\
Pointed {\it Chandra} Observation \\
~~RA (J2000) & 23$^h$ 58$^m$ 02$^s$.9 & \\
~~Dec (J2000) & -32$^{\circ}$ 34$'$ 01$''$.3 & \\
~~ObsID/SeqID & 3954/600328 & \\
~~Roll Angle (deg) & 30.2 & \\
~~Effective Exposure Time (sec) & 49094 & 
\enddata
\tablerefs{(1)  NED, (2) \citet{Tully88}, (3) This value has
been calculated based on the HI mass of NGC 7793 that was calculated by 
\citet{Carignan90} to be 6.8$\times$10$^8$ $M$$_{\rm{\odot}}$ for an
assumed distance to the galaxy of 3.38 Mpc. We have scaled this
value based on our different assumed distance to the galaxy. 
(4) \citet{Karachentsev03}.}
\end{deluxetable}

\clearpage


\clearpage

\begin{deluxetable}{lccccccc}
\rotate
\tabletypesize{\scriptsize}
\tablecaption{Properties of {\it Chandra}-Detected Sources in the NGC 7793 Field\label{Table2}}
\tablewidth{0pt}
\tablehead{
& & & \colhead{Net} & \colhead{Absorbed} & \colhead{Unbsorbed} & \colhead{Unabsorbed} & \\
& \colhead{R.A.} & \colhead{Decl.} & \colhead{Count} & \colhead{Flux} & 
\colhead{Flux} & \colhead{Luminosity} & \colhead{Significance}\\
\colhead{Source} & \colhead{(J2000.0)} & \colhead{(J2000.0)} &
\colhead{Rate (${\pm}$Error)} & \colhead{(${\times}$10$^{-15}$)} & \colhead{(${\times}$10$^{-15}$)} &
\colhead{(${\times}$10$^{37}$)} & \colhead{($\sigma$)} 
}
\startdata
CXOU J235743.8$-$323633 & 23 57 43.8 & $-$32 36 33.6 & 4.71(${\pm}$0.98)$\times$10$^{-4}$ & 4.78($\pm$0.99) & 5.02($\pm$1.10) & 0.92($\pm$0.20) & 4.6 \\
CXOU J235746.7$-$323607 & 23 57 46.7 & $-$32 36 07.4 & 5.50(${\pm}$0.33)$\times$10$^{-3}$ & 55.8($\pm$3.4)  & 57.8($\pm$3.5)  & 10.6($\pm$6.4)  & 18.5 \\
CXOU J235747.2$-$323523 & 23 57 47.2 & $-$32 35 23.8 & 4.79(${\pm}$0.98)$\times$10$^{-4}$ & 4.86($\pm$0.99) & 5.03($\pm$1.03) & 0.92($\pm$0.19) & 4.5 \\
CXOU J235748.6$-$323234 & 23 57 48.6 & $-$32 32 34.0 & 1.41(${\pm}$0.77)$\times$10$^{-4}$ & 1.43($\pm$0.79) & 1.48($\pm$0.82) & 0.27($\pm$0.15) & 2.9 \\
CXOU J235750.9$-$323728 & 23 57 50.9 & $-$32 37 28.5 & 3.24(${\pm}$0.26)$\times$10$^{-3}$ & 32.9($\pm$2.6)  & 34.1($\pm$2.7)  & 6.24($\pm$0.49) & 11.3 \\
CXOU J235750.9$-$323726 & 23 57 50.9 & $-$32 37 26.6 & 6.61(${\pm}$0.12)$\times$10$^{-2}$ & 671($\pm$12)  & 695($\pm$13)  & 127($\pm$2)  & 59.7 \\
CXOU J235752.7$-$323309 & 23 57 52.7 & $-$32 33 09.8 & 4.46(${\pm}$0.30)$\times$10$^{-3}$ & 45.3($\pm$3.1)  & 46.9($\pm$3.2)  & 8.58($\pm$0.59) & 23.6 \\
CXOU J235756.2$-$323136 & 23 57 56.2 & $-$32 31 36.5 & 3.08(${\pm}$1.02)$\times$10$^{-4}$ & 3.13($\pm$1.04) & 3.24($\pm$1.07) & 0.59($\pm$0.20) & 4.8  \\
CXOU J235756.3$-$323444 & 23 57 56.3 & $-$32 34 44.9 & 3.03(${\pm}$1.02)$\times$10$^{-4}$ & 3.07($\pm$1.04) & 3.18($\pm$1.08) & 0.58($\pm$0.20) & 6.4  \\
CXOU J235756.4$-$323559 & 23 57 56.4 & $-$32 35 59.8 & 1.31(${\pm}$0.16)$\times$10$^{-3}$ & 13.3($\pm$1.7)  & 13.8($\pm$1.7)  & 2.52($\pm$0.31) & 12.1 \\
CXOU J235759.8$-$323240 & 23 57 59.8 & $-$32 32 40.9 & 1.66(${\pm}$0.18)$\times$10$^{-3}$ & 16.8($\pm$1.9)  & 17.4($\pm$2.0)  & 3.18($\pm$0.37) & 14.9  \\
CXOU J235800.1$-$323325 & 23 58 00.1 & $-$32 33 25.4 & 9.70(${\pm}$1.41)$\times$10$^{-4}$ & 9.84($\pm$1.46) & 10.2($\pm$1.5)  & 1.87($\pm$0.27) & 7.4  \\
CXOU J235800.3$-$323455 & 23 58 00.3 & $-$32 34 55.0 & 1.09(${\pm}$0.15)$\times$10$^{-3}$ & 11.1($\pm$1.6)  & 11.5($\pm$1.6)  & 2.10($\pm$0.29) & 10.3 \\
CXOU J235800.7$-$323239 & 23 58 00.7 & $-$32 32 39.5 & 3.40(${\pm}$0.11)$\times$10$^{-4}$ & 3.45($\pm$1.08) & 3.57($\pm$1.12) & 0.65($\pm$0.20) & 5.5 \\
CXOU J235802.8$-$323614 & 23 58 02.8 & $-$32 36 14.3 & 2.76(${\pm}$0.24)$\times$10$^{-3}$ & 28.0($\pm$2.4)  & 29.0($\pm$2.5)  & 5.31($\pm$0.46) & 20.6 \\
CXOU J235803.5$-$323643 & 23 58 03.5 & $-$32 36 43.8 & 1.24(${\pm}$0.16)$\times$10$^{-3}$ & 12.6($\pm$1.6)  & 13.0($\pm$1.7)  & 2.38($\pm$0.31) & 10.9 \\
CXOU J235805.5$-$323250 & 23 58 05.5 & $-$32 32 50.8 & 2.04(${\pm}$0.87)$\times$10$^{-4}$ & 2.07($\pm$0.89) & 2.14($\pm$0.92) & 0.39($\pm$0.17) & 3.7  \\
CXOU J235806.6$-$323757 & 23 58 06.6 & $-$32 37 57.1 & 8.48(${\pm}$0.42)$\times$10$^{-3}$ & 86.0($\pm$4.3)  & 89.1($\pm$4.4)  & 16.3($\pm$0.8)  & 21.6  \\
CXOU J235807.8$-$323614 & 23 58 07.8 & $-$32 36 14.4 & 2.08(${\pm}$0.88)$\times$10$^{-4}$ & 2.11($\pm$0.89) & 2.19($\pm$0.92) & 0.40($\pm$0.17) & 3.9  \\
CXOU J235808.7$-$323403 & 23 58 08.7 & $-$32 34 03.7 & 5.11(${\pm}$0.32)$\times$10$^{-3}$ & 51.9($\pm$3.3)  & 53.7($\pm$3.4)  & 9.82($\pm$0.62) & 31.5 \\
CXOU J235809.6$-$323617 & 23 58 09.6 & $-$32 36 17.0 & 4.11(${\pm}$0.92)$\times$10$^{-4}$ & 4.17($\pm$0.93) & 4.32($\pm$0.96) & 0.79($\pm$0.18) & 5.7  \\
CXOU J235810.4$-$323357 & 23 58 10.4 & $-$32 33 57.6 & 2.47(${\pm}$0.22)$\times$10$^{-3}$ & 25.1($\pm$2.3)  & 26.0($\pm$2.4)  & 4.76($\pm$0.44) & 19.4  \\
\enddata


\tablecomments{Units of Right Ascension are hours, minutes and seconds
and units of Declination are degrees, arcminutes and arcseconds. Net
count rates are in units of counts s$^{-1}$ and refer to the number of
counts extracted above background in the 0.2-10.0 keV energy range
within the source extract region (for the entire observation with an
effective exposure time of 49094 seconds).  Fluxes (both absorbed and
unabsorbed) are in units of ergs cm$^{-2}$ s$^{-1}$ and have been
computed assuming a foreground column density $N$$_H$ =
1.15$\times$10$^{20}$ cm$^{-2}$ and adopting a power law model with a
photon index $\Gamma$=1.5. Unabsorbed luminosities 
are in units of ergs s$^{-1}$ and have been calculated assuming a
distance $d$ = 3.91 Mpc to NGC 7793.}


\end{deluxetable}

\clearpage

\begin{deluxetable}{lc}
\tabletypesize{\scriptsize}
\tablecaption{Multi-Wavelength Counterparts of {\it Chandra} Sources in the NGC 7793 
Field\label{Table3}}
\tablewidth{0pt}
\tablehead{
\colhead{Source} & \colhead{Counterparts} 
}
\startdata
CXOU J235743.8$-$323633 & D22 \\
CXOU J235746.7$-$323607 & P10, R3 \\
CXOU J235747.2$-$323523 & N7793-S11, D40 \\
CXOU J235748.6$-$323234 & P6, Star (USNO 0574$-$1250312), H18(?)\\
CXOU J235750.9$-$323728 & P13?\\
CXOU J235750.9$-$323726 & P13?\\
CXOU J235752.7$-$323309 & P7, Star (USNO 0574$-$1250339)\\
CXOU J235800.1$-$323325 & P8, N7793-S26 \\
CXOU J235802.8$-$323614 & P11?\\
CXOU J235803.5$-$323643 & P11?\\
CXOU J235808.7$-$323403 & P9 \\
\enddata

\tablecomments{The listed counterparts have been taken from the
  following references: H and D -- HII regions identifed by
  \citet{Hodge69} and \citet{Davoust80}, respectively; S --
  optically-identified SNRs identified by \citet{Blair97}; P -- X-ray
  sources detected by the $\it{ROSAT}$ PSPC and presented by RP99; R
  -- candidate radio SNRs identified by \citet{Pannuti02}. The
  identifications of stellar counterparts are made based on the USNO-B
  catalog \citep{Monet03}.  Sources not listed do not have any
  counterparts at other wavelengths.}

\end{deluxetable}

\clearpage

\begin{deluxetable}{lcccccccc}
\tabletypesize{\small}
\rotate
\tablecaption{Spectral Fits for the Brightest Sources\tablenotemark{a}\label{Table4}}
\tablewidth{0pt}
\tablehead{
& & \colhead{Energy}  & \colhead{kT or ${\Gamma}$} & \colhead{Col Density} & 
& \colhead{Gauss} & 
& \colhead{$\chi$$^2$/Degrees} \\
\colhead{Source} & \colhead{Model} & \colhead{Range} & \colhead{Value} & 
\colhead{N$_{\rm H}$} & Norm & Line Energy & EqW & \colhead{of Freedom} 
}
\startdata
CXOU J235746.7$-$323607 & Brems & 0.5--6.0 & 1.55$^{+2.82}_{-0.89}$ & 0.32$^{+0.66}_{-0.22}$ 
&  1.4($^{+6.5}_{-0.8}$)$\times$10$^{-5}$ & G1: 0.94f & 110$^{+75}_{-38}$ & 17.4/21=0.82\\
&        &          &                        &         &                                       & G2: 1.47f & 66$^{+70}_{-62}$    &  \\  
& DiskBB & 0.5--6.0 & 0.72$^{+0.47}_{-0.30}$ & $<$0.68 &  
4.4($^{+6.1}_{-3.8}$)$\times$10$^{-3}$ & G1: 0.94f & 109$^{+84}_{-12}$ & 16.7/21=0.80\\
&        &          &                        &         &                                       & G2: 1.47f & 65$^{+50}_{-48}$    &  \\  
& PowLaw & 0.5--6.0 & 2.87$^{+2.03}_{-0.80}$ & 0.47$^{+0.83}_{-0.17}$ & 
1.4($^{+7.2}_{-0.9}$)$\times$10$^{-5}$  & G1: 0.94f & 118$^{+90}_{-25}$ & 18.1/20=0.91\\
&        &          &                        &         &                                       & G2: 1.46f & $<$68             &  \\  
& PowLaw-f & 0.5--6.0 & 1.69$^{+0.51}_{-0.41}$ & 0.0115f & 
1.4($^{+7.2}_{-0.9}$)$\times$10$^{-5}$ & G1: 0.94f & 111$^{+74}_{-74}$ & 18.1/20=0.91\\
&        &          &                        &         &                                       & G2: 1.46f & 102$^{+76}_{-75}$    &  \\  
& APEC & 0.5--6.0 & 0.78$^{+0.18}_{-0.17}$ & 0.96$^{+0.21}_{-0.23}$ &
4.45($^{+2.17}_{-1.77}){\times}10^{-5}$ & $\dots$ & $\dots$ & 23.23/22=1.06 \\
CXOU J235750.9$-$323726 & Brems & 0.3--7.0 & $>$14 & 0.12$^{+0.07}_{-0.05}$ & 
6.4($^{+1.6}_{-1.1}$)$\times$10$^{-5}$ & G1: 0.59f & $<$48 & 66.8/65=1.03\\
&        &          &                       &                        &                                        & G2: 0.82f & 
42$^{+55}_{-39}$ &  \\  
&        &          &                       &                        &                                        & G3: 0.99f & $<$34 &  \\  
& DiskBB & 0.3--7.0 & 1.83$^{+0.16}_{-0.14}$ & 0.07$^{+0.03}_{-0.02}$ & 
6.0($^{+5.5}_{-3.2}$)$\times$10$^{-4}$  & G1: 0.82f & $<$132            & 66.3/65=1.02 \\
&        &          &                        &                       &                                        & G2: 0.82f & 
45$^{+50}_{-40}$ &  \\  
&        &          &                        &                       &                                        & G3: 0.99f & $<$30 &  \\  
& PowLaw & 0.3--7.0 & 1.4$^{+0.20}_{-0.18}$ & 0.17$^{+0.09}_{-0.07}$ &  
3.6($^{+0.6}_{-0.7}$)$\times$10$^{-5}$  & G1: 0.60f & $<$48            & 78.3/68=1.15\\
&        &          &                       &                        &                                        & G2: 0.82f & 
45$^{+63}_{-40}$ &  \\  
&        &          &                       &                        &                                        & G3: 0.99f & $<$69 &  \\  

CXOU J235806.6$-$323757 & Brems & 0.5--6.0 & 1.98$^{+1.24}_{-0.89}$ &
0.21$^{+0.21}_{-0.12}$ 
& 2.7($^{+2.7}_{-0.9}$)$\times$10$^{-5}$ & G1: 0.67f & 118$^{+170}_{-90}$ & 24.9/29=0.86 \\
                   &       &          &                        &                        &                                      & G2: 0.83f 
                   & 120$^{+95}_{-75}$ &   \\
& Brems-f & 0.5--6.0 & 6.42$^{+15.72}_{-3.08}$ & 0.0115f & 
2.7($^{+2.7}_{-0.9}$)$\times$10$^{-5}$ & G1: 0.67f & $<$51 & 24.9/29=0.86 \\
                     &                       &          &                        &                        &      & G2: 0.83f 
                     & 45$^{+27}_{-27}$ &   \\
& DiskBB & 0.5--6.0 & 0.83$^{+0.30}_{-0.22}$ & $<$0.08 & 
5.3($^{+13.4}_{-3.6}$)$\times$10$^{-3}$  & G1: 0.83f & 51$^{+39}_{-31}$ & 30.8/31=0.99\\
& PowLaw & 0.5--6.0 & 2.45$^{+0.65}_{-0.45}$ & 0.29$^{+0.17}_{-0.16}$ & 
2.5($^{+1.5}_{-0.8}$)$\times$10$^{-5}$  & G1: 0.67f & 122$^{+73}_{-42}$ & 23.9/31=0.77 \\
&        &          &                        &                        &                                       & G2: 0.83f 
& 118$^{+181}_{-75}$ &   \\
CXOU J235808.7$-$323403 & APEC & 0.5-2.5 & 0.83$^{+0.48}_{-0.65}$ &
0.16$^{+0.33}_{-0.08}$ & 
5.6($\pm$4.5)$\times$10$^{-4}$ & $\dots$ & $\dots$ & 16.5/15=1.10 \\
& APEC-f & 0.5-2.5 & 0.62$^{+0.32}_{-0.12}$ & 0.0115f & 1.7($\pm$1.4)$\times$10$^{-6}$
& $\dots$ & $\dots$ & 21.4/16=1.34 \\
& DiskBB & 0.5-2.5 & 0.20$^{+0.18}_{-0.09}$ & 0.28$^{+0.24}_{-0.18}$ & 1.8$\pm$1.4
& $\dots$ & $\dots$ & 19.2/15=1.28 \\

\enddata

\tablenotetext{a}{An ``f" after a number or label indicates that the
parameter was fixed at that value during the fit.  ``Energy Range" is
in units of keV.  Parameter values are as follows: Power Law (``PowLaw"): Photon
Index $\Gamma$ defined such that $E$$^{-\Gamma}$; Bremsstrahlung (``Brems"):
temperature $\it{kT}$ in keV and DiskBB (multi-color disk model)
temperature $\it{kT_{\rm{in}}}$ in keV. The normalizations are defined
as follows -- Power Law: photons/keV/cm$^{2}$/s at 1 keV;
Bremsstrahlung:
3.02$\times$10$^{-15}$/(4$\pi$$d$$^2$)~$\int$~$n$$_e$$n$$_i$dV where
$d$ is the distance to the source in cm and $n$$_e$ and $n$$_i$ are
the electron and ion densities respectively in cm$^{-3}$; DiskBB:
($R$$_{\rm{in}}$/$D$)$^2$~cos~$\theta$, where $R$$_{\rm{in}}$ is the
inner disk radius in km, $D$ is the distance to the source (in units
of 10 kpc) and $\theta$ is the angle of the disk.  The column density
$N$$_{\rm H}$ is in units of 10$^{22}$ cm$^{-2}$.  Finally, ``EqW" is the Gaussian
Equivalent Width in eV. In this table we are presenting the best fits derived from the
models that we used and not necessarily every model that we used. For details
of the spectral analysis of these sources, see Section \ref{LuminousXraySourcesSection}.}

\end{deluxetable}

\clearpage
                   
\begin{deluxetable}{lccccc}
\tabletypesize{\scriptsize}
\tablecaption{Quantile Values for Point Sources\label{Table5}}
\tablewidth{0pt}
\tiny
\tablehead{
\colhead{Source} & \colhead{$Q$$_{25}$} & \colhead{$Q$$_{50}$} & 
\colhead{$Q$$_{75}$} & \colhead{log($Q$$_{50}$/(1-$Q$$_{50}$))} & 
\colhead{3$Q$$_{25}$/$Q$$_{75}$}
}
\startdata
CXOU J235743.8-323633 & 0.096($\pm$0.004) & 0.132($\pm$0.011) & 0.280($\pm$0.018) & 
-0.819($\pm$0.118) & 1.029($\pm$0.148)\\ 
CXOU J235746.7-323607 & 0.077($\pm$0.001) & 0.129($\pm$0.001) & 0.223($\pm$0.005) & 
-0.828($\pm$0.062) & 1.036($\pm$0.078)\\ 
CXOU J235747.2-323523 & 0.051($\pm$0.004) & 0.061($\pm$0.003) & 0.079($\pm$0.002) & 
-1.187($\pm$0.061) & 1.951($\pm$0.101)\\ 
CXOU J235748.6-323234 & 0.028($\pm$0.067) & 0.059($\pm$0.038) & 0.122($\pm$0.014) & 
-1.204($\pm$0.409) & 0.700($\pm$0.238)\\ 
CXOU J235750.9-323728 & 0.085($\pm$0.000) & 0.141($\pm$0.001) & 0.269($\pm$0.001) & 
-0.785($\pm$0.028) & 0.945($\pm$0.034)\\ 
CXOU J235750.9-323726 & 0.082($\pm$0.000) & 0.132($\pm$0.000) & 0.258($\pm$0.000) & 
-0.816($\pm$0.018) & 0.951($\pm$0.021)\\ 
CXOU J235752.7-323309 & 0.080($\pm$0.004) & 0.126($\pm$0.001) & 0.228($\pm$0.008) & 
-0.841($\pm$0.089) & 1.054($\pm$0.112)\\ 
CXOU J235756.2-323136 & 0.089($\pm$0.011) & 0.107($\pm$0.030) & 0.197($\pm$0.143) & 
-0.923($\pm$0.311) & 1.355($\pm$0.456)\\ 
CXOU J235756.3-323444 & 0.088($\pm$0.027) & 0.129($\pm$0.044) & 0.201($\pm$0.115) & 
-0.829($\pm$0.272) & 1.314($\pm$0.431)\\ 
CXOU J235756.4-323559 & 0.073($\pm$0.010) & 0.108($\pm$0.004) & 0.170($\pm$0.014) & 
-0.918($\pm$0.125) & 1.286($\pm$0.175)\\ 
CXOU J235759.8-323240 & 0.065($\pm$0.007) & 0.110($\pm$0.004) & 0.183($\pm$0.016) & 
-0.908($\pm$0.134) & 1.064($\pm$0.157)\\ 
CXOU J235800.1-323325 & 0.049($\pm$0.003) & 0.066($\pm$0.001) & 0.081($\pm$0.001) & 
-1.154($\pm$0.059) & 1.834($\pm$0.093)\\ 
CXOU J235800.3-323455 & 0.046($\pm$0.003) & 0.057($\pm$0.002) & 0.074($\pm$0.002) & 
-1.218($\pm$0.063) & 1.863($\pm$0.096)\\ 
CXOU J235800.7-323239 & 0.076($\pm$0.034) & 0.110($\pm$0.024) & 0.176($\pm$0.274) & 
-0.908($\pm$0.444) & 1.285($\pm$0.629)\\ 
CXOU J235802.8-323614 & 0.060($\pm$0.002) & 0.099($\pm$0.006) & 0.157($\pm$0.005) & 
-0.960($\pm$0.079) & 1.142($\pm$0.094)\\ 
CXOU J235803.5-323643 & 0.073($\pm$0.007) & 0.111($\pm$0.010) & 0.186($\pm$0.045) & 
-0.904($\pm$0.192) & 1.169($\pm$0.248)\\ 
CXOU J235805.5-323250 & 0.081($\pm$0.098) & 0.112($\pm$1.079) & 0.388($\pm$0.671) & 
-0.900($\pm$0.998) & 0.626($\pm$0.694)\\ 
CXOU J235806.6-323757 & 0.068($\pm$0.000) & 0.111($\pm$0.001) & 0.179($\pm$0.001) & 
-0.904($\pm$0.035) & 1.139($\pm$0.044)\\ 
CXOU J235807.8-323614 & 0.069($\pm$0.035) & 0.094($\pm$0.022) & 0.139($\pm$0.346) & 
-0.983($\pm$0.496) & 1.497($\pm$0.755)\\ 
CXOU J235808.7-323403 & 0.057($\pm$0.001) & 0.087($\pm$0.002) & 0.137($\pm$0.003) & 
-1.024($\pm$0.057) & 1.240($\pm$0.069)\\ 
CXOU J235809.6-323617 & 0.055($\pm$0.012) & 0.081($\pm$0.046) & 0.174($\pm$0.289) & 
-1.056($\pm$0.593) & 0.954($\pm$0.536)\\ 
CXOU J235810.4-323357 & 0.079($\pm$0.007) & 0.125($\pm$0.002) & 0.185($\pm$0.025) & 
-0.846($\pm$0.134) & 1.277($\pm$0.202) \\
\enddata
\tablecomments{$Q$$_{25}$, $Q$$_{50}$ and $Q$$_{75}$ are all defined in Section
\ref{QuantileSection}.}
\end{deluxetable}

\clearpage

\begin{deluxetable}{lcccccc}
\tabletypesize{\scriptsize}
\tablecaption{Unabsorbed Luminosities of Common Discrete X-ray Sources Detected
by {\it ROSAT} and {\it Chandra}\tablenotemark{a}\label{Table6}}
\tablewidth{0pt}
\tablehead{
\colhead{\it{ROSAT}} & \colhead{Unabsorbed} & \colhead{Unabsorbed} 
& \colhead{{\it Chandra}} & \colhead{Unabsorbed}\\
\colhead{Source} & \colhead{$L$$_{\rm{Einstein}}$} & 
\colhead{$L$$_{\rm{PSPC}}$} & \colhead{Source} & \colhead{$L$$_{\rm{Chandra}}$} 
}
\startdata
P10 & \nodata & 1.35($\pm$0.14)$\times$10$^{38}$ & CXOU J235746.7$-$323607 & 
4.79($\pm$0.28)$\times$10$^{37}$ \\
P6 & \nodata & 3.14($\pm$0.71)$\times$10$^{37}$ & CXOU J235748.6$-$323234 & 
1.23($\pm$0.70)$\times$10$^{36}$ \\
P7 & \nodata & 4.20($\pm$0.88)$\times$10$^{37}$ & CXOU J235752.7$-$323309 & 
3.88($\pm$0.26)$\times$10$^{37}$  \\
P8 & \nodata & 5.43($\pm$0.88)$\times$10$^{37}$ & CXOU J235800.1$-$323325 & 
8.43($\pm$1.22)$\times$10$^{36}$ \\
P9 & \nodata & 8.57($\pm$1.05)$\times$10$^{37}$ & CXOU J235808.7$-$323403 & 
4.45($\pm$0.28)$\times$10$^{37}$  \\
P13 & 7.43($\pm$1.49)$\times$10$^{38}$ & 7.62($\pm$0.25)$\times$10$^{38}$ & 
CXOU J235750.9$-$323726 & 5.74($\pm$0.11)$\times$10$^{38}$   \\
& & & CXOU J235750.9$-$323728 & 2.82($\pm$0.22)$\times$10$^{37}$ \\
P11 & \nodata & 3.14($\pm$0.71)$\times$10$^{37}$ & CXOU J235802.8$-$323614 & 
2.40($\pm$0.20)$\times$10$^{37}$ \\ 
& & & CXOU J235803.5$-$323643 & 1.08($\pm$0.14)$\times$10$^{37}$  
\enddata
\tablenotetext{a}{The luminosities have been calculated by converting count
rates (over the range of 0.12-2.48 keV) to fluxes using a power-law model
(with a photon index $\Gamma$=1.5) and assuming a column density 
$N$$_H$=1.15$\times$10$^{20}$ cm$^{-2}$. The units of the luminosities are
ergs s$^{-1}$.}
\end{deluxetable}

\clearpage

\begin{deluxetable}{lcccccc}
\tabletypesize{\scriptsize}
\tablecaption{Spectral Fits to Diffuse Emission 
Spectrum\tablenotemark{a}\label{diffuse-fit}}
\tablewidth{0pt}
\tablehead{
& \colhead{$N$$_{\rm H}$} & & &  \colhead{Flux} & \colhead{Unabsorbed} 
& \colhead{$\chi$$^2$/Degrees}\\
\colhead{Model} &  
\colhead{(10$^{22}$ cm$^{-2}$)} & \colhead{Parameter\tablenotemark{b}} & 
\colhead{Norm} & \colhead{or EqW\tablenotemark{c}} & \colhead{Flux} &
\colhead{of Freedom}
}
\startdata
APEC        & 0.368$^{+0.051}_{-0.052}$ & 0.193$^{+0.013}_{-0.009}$ & 
1.0$\pm$0.1$\times$10$^{-3}$ & 5.4$\times$10$^{-13}$  & 17.4$\times$10$^{-13}$ 
& 763.58/730=1.05\\
 + Power Law & ${\cdots}$ & 1.15$^{+0.18}_{-0.17}$  & 6.1$\pm$0.6$\times$10$^{-5}$ & 
0.8$\times$10$^{-13}$ & 4.9$\times$10$^{-13}$ & ${\cdots}$ \\
VAPEC  & 0.015f     & 0.253$^{+0.018}_{-0.015}$  & 6.1$^{+0.4}_{-0.8}$$\times$10$^{-5}$ 
& 5.4$\times$10$^{-13}$ & 5.5$\times$10$^{-13}$ & 806.29/731=1.10\\
 + Ne    & ${\cdots}$ & 2.23$^{+0.74}_{-0.66}$ & ${\cdots}$ & ${\cdots}$ & ${\cdots}$ &
 ${\cdots}$\\
 + Power Law & ${\cdots}$ & 0.86$^{+0.15}_{-0.14}$ & 4.0$^{+0.4}_{-0.3}$$\times$10$^{-5}$ 
 & 4.5$\times$10$^{-13}$ & 4.6$\times$10$^{-13}$ & ${\cdots}$ \\
APEC-1  & 0.015f     & 0.218$^{+0.026}_{-0.016}$ & 5.3${\pm}$0.7$\times$10$^{-5}$ & 
0.7$\times$10$^{-13}$ & 0.8$\times$10$^{-13}$ & 764.08/727=1.05\\
 + APEC-2  & ${\cdots}$ & 0.547$^{+0.106}_{-0.117}$  & 1.7${\pm}$0.4$\times$10$^{-5}$ 
 & 0.4$\times$10$^{-13}$ & 0.5$\times$10$^{-13}$ & ${\cdots}$ \\
 + Power Law & ${\cdots}$ & 0.75$\pm$0.14  & 3.8${\pm}$0.3$\times$10$^{-5}$ & 
1.0$\times$10$^{-13}$ & 1.1$\times$10$^{-13}$ & ${\cdots}$ \\
Multiple Gaussians & 0.226$^{+0.195}_{-0.093}$ & ${\cdots}$ & ${\cdots}$ & 
5.5$\times$10$^{-13}$ & 7.3$\times$10$^{-13}$ & 784.78/728=1.08\\
 + Gaussian-1  & ${\cdots}$ & 0.603${\pm}$0.015 & 2.5${\pm}$0.4$\times$10$^{-4}$ 
 & ~49 & ${\cdots}$ & ${\cdots}$ \\ 
 + Gaussian-2  & ${\cdots}$ & 0.751${\pm}$0.009 & 5.9${\pm}$0.9$\times$10$^{-5}$ 
 & 101 & ${\cdots}$ & ${\cdots}$ \\
 + Gaussian-3  & ${\cdots}$ & 0.896${\pm}$0.013 & 3.0${\pm}$0.5$\times$10$^{-5}$ & ~85 
 & ${\cdots}$ & ${\cdots}$\\
 + Power Law & ${\cdots}$ & 1.36$^{+0.17}_{-0.15}$ & 7.4${\pm}$0.6$\times$10$^{-5}$ 
 & 4.5$\times$10$^{-13}$ & 4.5$\times$10$^{-13}$ & ${\cdots}$ \\
Multiple Gaussians & 0.015f  & ${\cdots}$ & ${\cdots}$ & 1.9$\times$10$^{-13}$ & 
2.0$\times$10$^{-13}$ & 808.84/730=1.11 \\
 + Gaussian-1  & ${\cdots}$ & 0.608${\pm}$0.011 & 4.3${\pm}$0.8$\times$10$^{-5}$ & 39 
 & ${\cdots}$ & ${\cdots}$ \\
 + Gaussian-2  & ${\cdots}$ & 0.753${\pm}$0.015 & 1.8${\pm}$0.3$\times$10$^{-5}$ & 74 
 & ${\cdots}$ & ${\cdots}$ \\
 + Gaussian-3  & ${\cdots}$ & 0.890${\pm}$0.020 & 1.3${\pm}$0.2$\times$10$^{-5}$ & 68 & 
 ${\cdots}$ & ${\cdots}$ \\
 + Power Law & ${\cdots}$ & 1.06${\pm}$0.12 & 5.1${\pm}$0.3$\times$10$^{-5}$ 
 & 1.2$\times$10$^{-13}$ & 1.3$\times$10$^{-13}$ & ${\cdots}$ \\
\enddata
\tablenotetext{a}{See Section
\ref{DiffuseSection} for a detailed description of these spectral fits. Also note that an ``f"
after a number indicates that the parameter was fixed at that value during the fit.}
\tablenotetext{b}{'Parameter' is defined as follows for the various
model components: for APEC, kT (keV); for the power law, the power law index
${\Gamma}$; for a specific element, the abundance; for a Gaussian, the
energy of the line center (keV).}
\tablenotetext{c}{Flux units = ergs s$^{-1}$ cm$^{-2}$ in
0.5-2 keV band; EqW = Equivalent width in eV. The flux in the background
in the 0.5-2 keV band is 3.4${\times}$10$^{-13}$ ergs s$^{-1}$ cm$^{-2}$; this
flux has been removed from the computed model fluxes for the diffuse emission
components.}
\end{deluxetable}

\clearpage

\begin{deluxetable}{cccc}
\tablecaption{Comparison of Inferred Temperatures to Fits of the Diffuse Emission of
Nearby Galaxies\tablenotemark{a}\label{diffuse-table}}
\tablewidth{0pt}
\tablehead{
& \colhead{$kT$$_1$} & \colhead{$kT$$_2$}\\  
\colhead{Galaxy} & \colhead{(keV)} & \colhead{(keV)} & \colhead{Reference}
}
\startdata
NGC 2403 & 0.18$\pm$0.03 & 0.73$\pm$0.07 & \citet{Fraternali02} \\
NGC 3184 & 0.13$^{+0.03}_{-0.03}$ & 0.43$^{+0.25}_{-0.22}$ & \citet{Doane04} \\
NGC 6946 & 0.25$^{+0.05}_{-0.02}$ & 0.71$^{+0.10}_{-0.08}$ & \citet{Schlegel03B} \\
NGC 7793 & 0.253$^{+0.018}_{-0.015}$ & $\cdots$ & This paper \\
\enddata
\tablenotetext{a}{See Section \ref{DiffuseSection}.}
\end{deluxetable}

\clearpage

\begin{deluxetable}{lcccc}
\rotate
\tabletypesize{\scriptsize}
\tablecaption{Comparison of Slopes of Luminosity Functions and Star Formation Rates
of Nearby Galaxies\tablenotemark{a}\label{lumfunction-table}}
\tablewidth{0pt}
\tablehead{
& \colhead{Slope} & & \colhead{Star Formation Rate} \\
\colhead{Galaxy} & \colhead{$\Gamma$} & \colhead{Reference} & \colhead{($M$$_{\odot}$
 yr$^{-1}$)} & \colhead{Reference}
 }
 \startdata
NGC 2403 & $-$0.59 & (1) & 0.30\tablenotemark{b} & (2) \\
M83 (NGC 5236) & $-$1.38$\pm$0.28 & (3) & 2.76 & (4) \\
NGC 6946 & $-$0.64 & (5) & 4 & (6) \\
NGC 7793\tablenotemark{c} & $\sim$$-$0.62$\pm$0.2, $\sim$$-$0.65$\pm$0.11 & (7) & 
0.24 & (8) \\
IC 5332 & $-$1.30$\pm$0.31 & (3) & 0.08 & (9) 
\enddata
\tablecomments{References -- (1) \citet{Schlegel03A}, (2) \citet{Sivan90}, 
(3) \citet{Kilgard02}, (4) \citet{Dopita10}, (5) \citet{Holt03}, (6) \citet{Sauty98},
(7) This paper, (8) \citet{Ferguson96}, (9) \citet{Meurer06}.}
\tablenotetext{a}{See Section \ref{LFSection}.}
\tablenotetext{b}{To estimate the star formation rate for NGC 2403, we took the
integrated H$\alpha$ luminosity of this galaxy as estimated by \citet{Sivan90}
($L$$_{H\alpha}$ = 3.4$\times$10$^{40}$ ergs sec$^{-1}$) and calculated
a star formation rate using the relation derived by \citet{Kennicutt83} (that is,
the total star formation rate in units of $M$$_{\odot}$ yr$^{-1}$ is expressed as
SFR(total) = $L$$_{H\alpha}$ / 1.12$\times$10$^{41}$ ergs sec$^{-1}$).} 
\tablenotetext{c}{As described in Section\ref{LFSection}, a slope of 
$\Gamma$$\sim$$-$0.62$\pm$0.2 was obtained for a fit to the complete function (though
the residuals are large) while a slope of $\Gamma$$\sim$$-$0.65$\pm$0.11 was obtained
for a fit that excluded the ULX.}
\end{deluxetable}

\end{document}